\documentclass[manuscript,screen,nonacm]{acmart}

\AtBeginDocument{%
  \providecommand\BibTeX{{%
    \normalfont B\kern-0.5em{\scshape i\kern-0.25em b}\kern-0.8em\TeX}}}

\setcopyright{acmlicensed}
\acmJournal{TORS}
\acmYear{2023} \acmVolume{1} \acmNumber{1} \acmArticle{1} \acmMonth{1} \acmPrice{15.00}\acmDOI{10.1145/3606950}

\usepackage{caption}
\usepackage{subcaption}
\usepackage{multirow}

\begin{document}

\title{Recommending Target Actions Outside Sessions in the Data-poor Insurance Domain}

\author{Simone Borg Bruun}
\email{simoneborgbruun@di.ku.dk}
\orcid{0000-0003-1619-4076}
\affiliation{%
  \institution{Department of Computer Science, University of Copenhagen}
  \streetaddress{Universitetsparken 1}
  \postcode{2100}
  \city{Copenhagen}
  \country{Denmark}
}

\author{Christina Lioma}
\email{c.lioma@di.ku.dk}
\orcid{0000-0003-2600-2701}
\affiliation{%
  \institution{Department of Computer Science, University of Copenhagen}
  \streetaddress{Universitetsparken 1}
  \postcode{2100}
  \city{Copenhagen}
  \country{Denmark}
}

\author{Maria Maistro}
\email{mm@di.ku.dk}
\orcid{0000-0002-7001-4817}
\affiliation{%
  \institution{Department of Computer Science, University of Copenhagen}
  \streetaddress{Universitetsparken 1}
  \postcode{2100}
  \city{Copenhagen}
  \country{Denmark}
}

\authorsaddresses{}

\renewcommand{\shortauthors}{Bruun et al.}

\begin{abstract}
Providing personalized recommendations for insurance products is particularly challenging due to the intrinsic and distinctive features of the insurance domain. First, unlike more traditional domains like retail, movie etc., a large amount of user feedback is not available and the item catalog is smaller. Second, due to the higher complexity of products, the majority of users still prefer to complete their purchases over the phone instead of online. We present different recommender models to address such data scarcity in the insurance domain. We use recurrent neural networks with 3 different types of loss functions and architectures (cross-entropy, censored Weibull, attention). Our models cope with data scarcity by learning from multiple sessions and different types of user actions. Moreover, differently from previous session-based models, our models learn to predict a target action that does not happen within the session. Our models outperform state-of-the-art baselines on a real-world insurance dataset, with ca. 44K users, 16 items, 54K purchases and 117K sessions. Moreover, combining our models with demographic data boosts the performance. Analysis shows that considering multiple sessions and several types of actions are both beneficial for the models, and that our models are not unfair with respect to age, gender and income.
\end{abstract}




\maketitle

\section{Introduction}
We present the problem of providing automatic personalised recommendations in the insurance purchasing domain. That is to recommend insurances for individuals such as home insurance, car insurance and accident insurance that can help customers of an insurance company continuously adjust their insurances to suit their needs. The following reasons make this problem particularly challenging: (i) the item catalog is small (only a few types of insurance products are available for purchase), unlike the common scenarios of e-commerce or movie recommendation, where the item catalog is very large; (ii) insurance products are purchased less often and tend to last longer, than items commonly bought in common recommendation scenarios; (iii) most users prefer to complete their purchases of insurance products in a telephone conversation with a human insurance agent who can interactively address their concerns, instead of fully and exclusively online. Collectively, the above reasons result in a limited amount of user interactions and feedback on the items available for recommendation, making the problem of automatically learning recommendations non-trivial.

To address this problem, we present three different architectures of recurrent neural network recommender models, which deal with the data scarcity problem outlined above in two ways. Firstly, they learn user preferences not only from single sessions and from user actions that are associated with an item, but from multiple past user sessions and several different types of user actions that are not necessarily associated with an item. 
Secondly, unlike existing session-based and session-aware recommender models, our models learn to predict a target action (purchase of an insurance item) that does not happen within the input session (because the purchase happens on the telephone conversation with a human insurance agent, for instance).

We show experimentally that these two features of our models make them outperform state-of-the-art recommender baselines on a real-world insurance dataset of ca.~44K users, 16 items, 54K purchases and 117K sessions (which is freely available to the research community).

This paper is an extended version of the conference full paper~\cite{BruunEtAl2022} accepted at RecSys 2022. We \textbf{contribute} by extending the previous work with (1) two new approaches to design the loss function and architecture of the proposed cross-sessions recommender system; (2) an evaluation of the novel approaches against the state-of-the-art and the original cross-session recommender system; (3) additional analysis (e.g., fairness analysis, analysis of the dependency on the temporal threshold); and (4) more details on the exploited dataset.

\section{Related Work}
We present prior work on insurance recommendation (Section~\ref{subsec:related_insurance}) and session-based recommendation (Section~\ref{subsec:related_session}).

\subsection{Insurance Recommendation}
\label{subsec:related_insurance}

There is not much prior work on insurance recommendations. In principle, knowledge-based recommendations should work for this task by mining highly personalised user information from user interactions \cite{HuiZZWN22}, but to our knowledge, no prior work reports this. Instead, most prior work on insurance recommendations supplements the small volume of user feedback with user demographics. We overview these next.

\citet{Xu2014} cluster users based on their demographics and make association rule analysis within each cluster on the users’ set of purchased items. They extract recommendations directly from the association rules. \citet{Mitra2014} estimate user similarity based on demographic attributes using a similarity measure (e.g., cosine similarity). Then they make recommendations to a user based on the feedback on items by the top-N similar users. \citet{Qazi2017, Qazi2020} train a Bayesian network with user demographics and previously purchased items as input features, aiming to predict the last purchased item of a user. They also train a feed-forward neural network to give recommendations to potential users when only external marketing data is available. All the above methods outperform standard RS approaches, such as matrix factorization and association rule mining solely applied to the feedback data. In addition, these methods are not susceptible to cold start issues when recommending items to users with no previous feedback on items~\cite{10.1145/3397271.3401060}. However, the above methods assume that the preference dynamics are homogeneous within demographic segments, which is not necessarily true. Our model addresses this by accommodating individual changes in user preferences through the use of sessions generated by the individual user.

\citet{Bi2020} present a cross-domain approach for insurance recommendation. They use knowledge from an e-commerce domain (clothes, skincare products, fruits, electronics products, etc.) to learn better recommendations in the insurance domain when data is sparse. They employ a Gated Recurrent Unit (GRU)~\cite{ChoEtAl2014} to model sequential dependencies in the e-commerce domain. Our model differs from this approach: we use user sessions directly from the insurance target domain instead of another source domain like an e-commerce website with clothes etc., thereby not having the need for overlapping users between the target domain and a source domain. 
Moreover, we cannot use the approach of \cite{Bi2020} in our work because it is not session-based; it is based on users' long-term preferences in both the e-commerce and the insurance domain.

\subsection{Session-based Recommender Systems}
\label{subsec:related_session}

Session-based RSs capture the user's short-term preferences in a session~\cite{FangEtAl2020}, often using item-to-item recommendations~\cite{Davidson2010,Linden2003}: similarities between items are computed based on the session data such that items that often co-occur and/or co-interact in a session fetch a high similarity. Such item similarities are then used during a session to recommend the most similar items to the item that the user currently interacts with. This approach only considers the last user interaction; it ignores information on past interactions even in the same session. Moreover, this approach requires the target action to happen within sessions, thus this approach cannot be applied to our task.

An extension of the item-to-item approach is session-based clustering, which considers all user interactions in the session. Sessions are then clustered in various ways, for instance as Markov chains~\cite{conf/edm/HansenHHAL17}, or using K-Nearest Neighbors (SKNN)~\cite{JannachAndLudewig2017,HuEtAl2020}, which computes similarities between entire sessions using a similarity measure (e.g., cosine similarity). The recommendations are then based on selecting items that appeared in the most similar past session. This approach does not take into account the order of the input sequence.

 A sequential extension to the SKNN method is Vector Multiplication SKNN~\cite{LudewigAndJannach2018}, which rewards the most recent user interactions of a session when computing the similarities. Another extension is Sequence and Time Aware Neighborhood~\cite{GargEtAL2019}, which considers the position of an item in the current session, the recency of a past session with respect to the current session, and the position of a recommendable item in a neighbouring session. The latter model depends on the target actions occurring within sessions, and so does not fit our task where the target actions occur outside the session.

Neural session-based methods have also been used. A popular one is GRU4REC~\cite{HidasiEtAl2016,HidasiAndKaratzoglou2018}, which models user sessions with a GRU in order to predict the probability of the subsequent interaction given the sequence of previous interactions. Neural Attentive Session-based Recommendation~\cite{LiEtAl2017} extends GRU4REC with an attention mechanism that captures the user’s main purpose in the current session. Graph Neural Networks have also been used ~\cite{WuEtAl2019} to model session sequences as graph structures, thereby capturing transitions of items and generating item embedding vectors correspondingly.

The above methods use only the single ongoing session of a user, whereas we use multiple past sessions of a user. Methods that use past user sessions when predicting the next interaction for the current session have been proposed~\cite{Quadrana2017, Ruocco2017, Ying2018, Hu2018, Phuong2019}. However, a comprehensive empirical study of these methods~\cite{Latifi2021} shows that they did not improve over heuristic extensions of existing session-based algorithms that for instance extend the current session with previous sessions or boost the scores of items previously interacted with. Unlike all the above methods, we use different types of actions, not only with items (see Section~\ref{subsec:dataset}).

Although session-based RSs provide temporal context for the recommendations in terms of the sequential order of interactions, they do not account for the actual time of interactions. Other types of RSs exploit time as a contextual variable in the learning of recommendations~\cite{Stormer2007, Brenner2010}. However, these methods use the specific property of time at the level of specificity of the hour, day, week, month or season to learn the recommendations. Other RSs integrate methods from survival analysis to predict users' return time to a service~\cite{Du2015, Jing2017} that for instance can help in planning advertising inventories. Our approach to account for time in the RS differs from these, since we do not focus on a single return time to estimate when a user comes back, rather we model times for each item to be used for the ranking task.

\section{Approach}
In this section, we present the problem formulation (Section~\ref{subsec:problem}) and how we address it with our approach (Section~\ref{subsec:approach}).

\subsection{Problem Formalization}
\label{subsec:problem}

\begin{figure}[tb]
    \centering
    \includegraphics[width=0.6\textwidth]{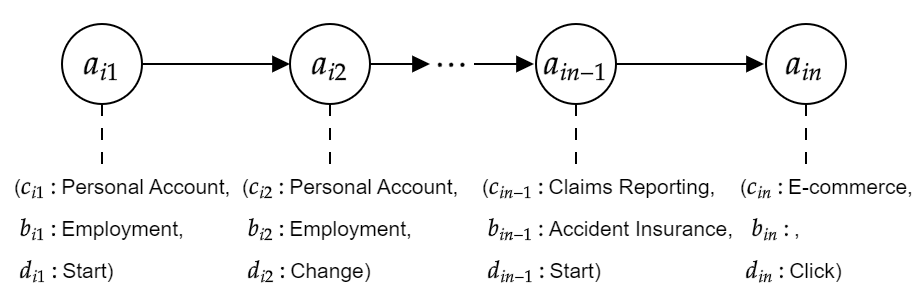}
    \caption{Example of a session on the insurance website. A session is a list of 3-tuple actions ordered by time.}
    \label{fig:insurance session}
\end{figure}

The task of our cross-sessions RS is to predict what items a user will buy based on the user's past sessions. Compared to the traditional task of a session-based RS, which is to predict the next action in the session based on the actions so far, our task differs because: (1) the target action (i.e., purchase) occurs outside the session; (2) the user may have multiple sessions leading to the purchase; (3) the sessions consist of many different actions, not only actions with items.
We extend the notation in~\citet{FangEtAl2020} to accommodate these differences when formulating the problem.

A session, $s_i$, is a sequence of user actions, $\lbrace a_{i1},a_{i2},a_{i3},...,a_{in} \rbrace$, on the website. An action, $a_{ij}$, is represented by the 3-tuple $a_{ij} = (c_{ij}, b_{ij}, d_{ij})$, where:
\begin{itemize}
    \item $c_{ij}$: action section, refers to the section of the website in which the user interacts;
    \item $b_{ij}$: action object, refers to the object on the website that the user interacts with; and
    \item $d_{ij}$: action type, refers to the way that the user interacts.
\end{itemize}
Figure~\ref{fig:insurance session} illustrates a session where a user interacts with employment (object $b_{i1}$) at the personal account section of the website (section $c_{i1}$) by starting it (action type $d_{i1}$). Then, the user changes ($d_{i2}$) the employment ($b_{i2}$) still at the personal account section ($c_{i1}$).
Section~\ref{subsec:dataset} and Table~\ref{tab:actions} present the different sections, objects and action types.

We represent users' past sessions in lists ordered by time. We do not include all historical sessions of a user as we assume that only recent sessions are relevant to the current task. We use an inactivity threshold, $t$, to define recent sessions, and define two sessions to belong to the same task if there is no longer than the threshold $t$ between them. We describe how we estimate $t$ from observed data in Section~\ref{subsec:dataset}.
The problem is to learn a function, $f$, of a user's session sequence that estimates the probability of the user to buy each item $k$ after the last session $s_m$:
\begin{equation}
    f(s_1,s_2,s_3,...,s_m) = (\hat{p}_1,\hat{p}_2,\hat{p}_3,...,\hat{p}_K),
    \label{eq:task}
\end{equation}
where each element in $\lbrace s_1,s_2,s_3,...,s_m \rbrace$ is a session (defined as a sequence of actions as explained above), $\hat{p}_k$ is the predicted probability that item $k$ will be bought by the user, and $K$ is the total number of items in the whole dataset.

\subsection{Proposed Approach}
\label{subsec:approach}

\begin{figure*}[tb]
\centering
\begin{minipage}[t]{.22\textwidth}
    \includegraphics[width=\linewidth]{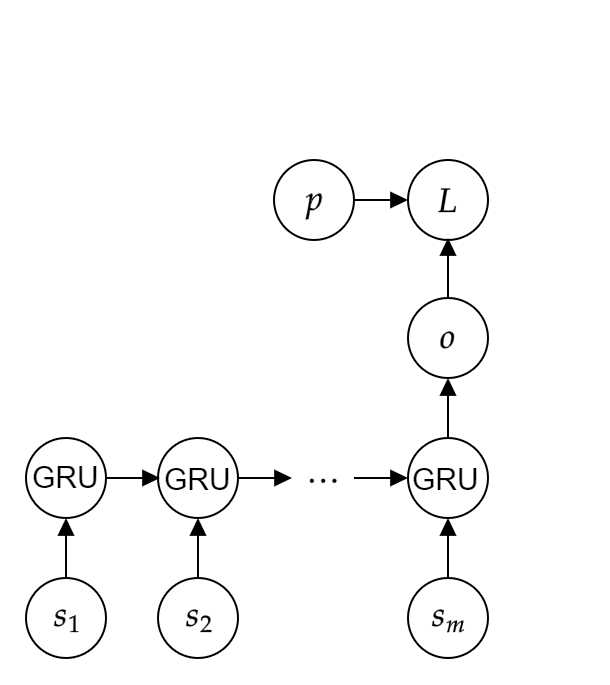}
    \caption{Architecture of cross-sessions encode.}
    \label{fig:architecture_encode}
\end{minipage}%
\begin{minipage}[t]{.44\textwidth}
    \includegraphics[width=\linewidth]{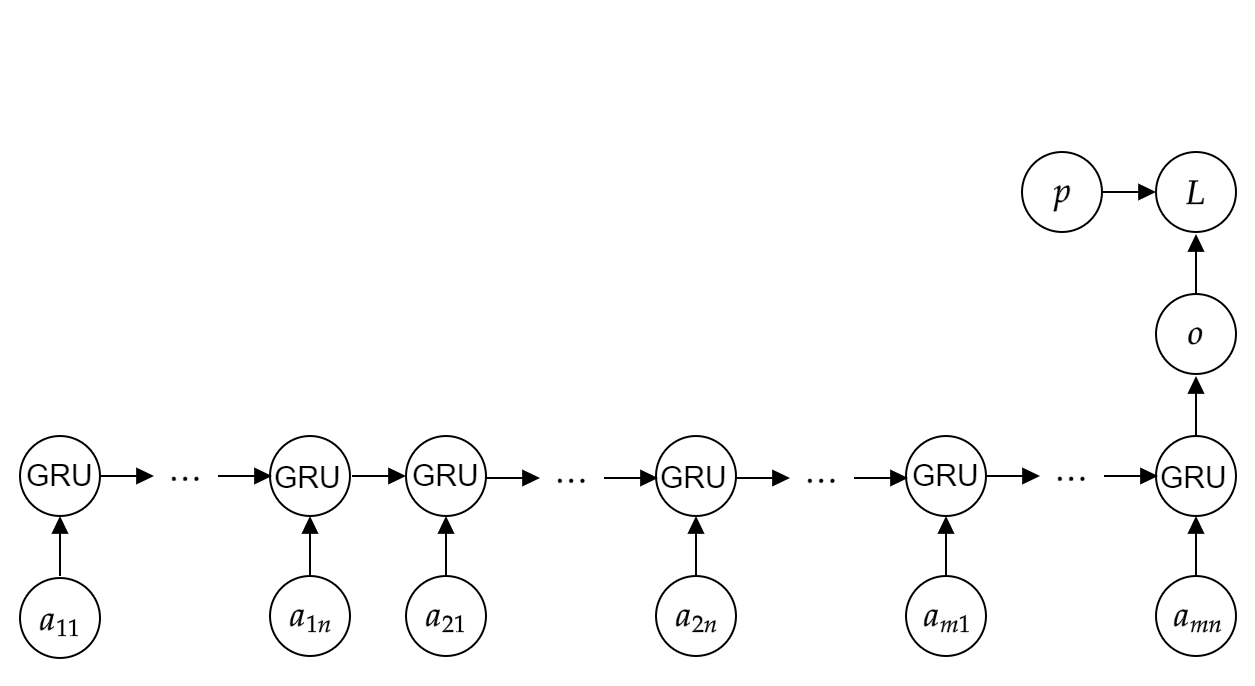}
    \caption{Architecture of cross-sessions concat.}
    \label{fig:architecture_concat}
\end{minipage}%
\begin{minipage}[t]{.33\textwidth}
    \includegraphics[width=\linewidth]{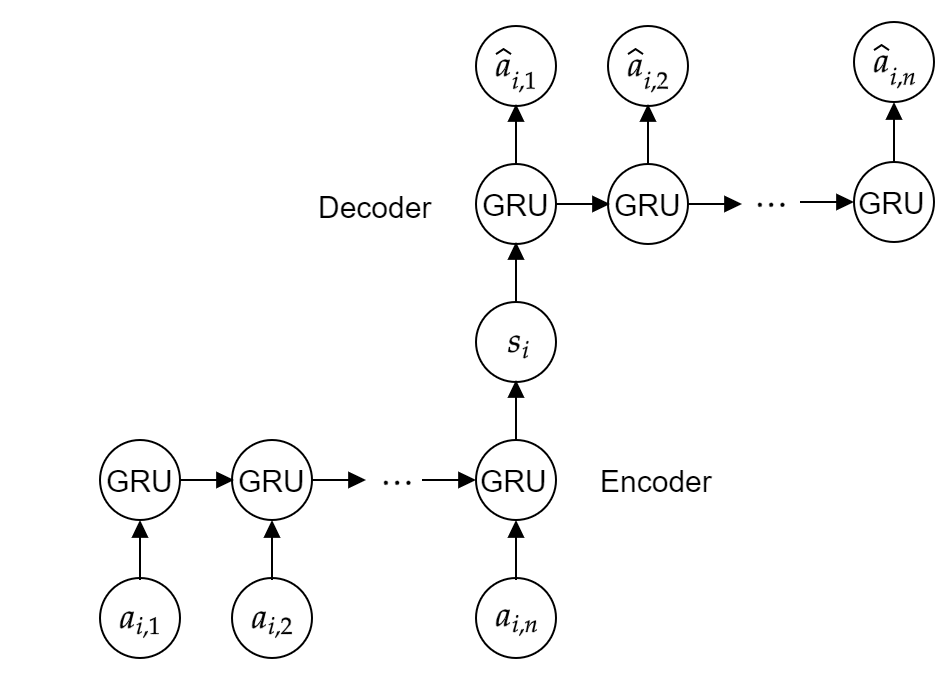}
    \caption{Architecture of cross-sessions auto.}
    \label{fig:architecture_auto}
\end{minipage}
\end{figure*}

Our model for learning the above function is based on GRU4REC~\cite{HidasiEtAl2016}: an RNN with a single GRU~\cite{ChoEtAl2014} layer that models user interactions with items in single user sessions. The RNN takes as input the ordered sequence of items interacted with in the session, and outputs for every time step the likelihood of each item to be the item that the user interacts with next. Our cross-sessions RS extends GRU4REC by: (1) taking multiple sessions of each user as input, as in Eq.~\eqref{eq:task}; (2) using various types of input actions that are not always associated with items; (3) predicting what items the user will buy after the last time step as opposed to predicting the next interaction for every time step in the sequence.

\paragraph{Model input} Next, we propose three different ways of passing the input sessions through the RNN.

In the first way, which we call \emph{Cross-sessions Encode} (see Figure~\ref{fig:architecture_encode}), we encode a session by aggregating the actions in the session with a maximum pooling operation:
\begin{equation}
    s_i = \text{max}_{element}(a_{i1},a_{i2},a_{i3},...,a_{in}),
    \label{eqn:encoding}
\end{equation}
where $a_{ij}$ is the binarized vector indicating the presence of an action section, action object and action type performed by a user at time step $j$ in session $i$, and $\max_{element}(\cdot)$ is a function that takes the element-wise maximum of vectors. Then, for every time step $i$ in the sequence of a user's sessions, an RNN with a single GRU layer computes the hidden state as follows:
\begin{align}
  \begin{aligned}
   h_i &= (1-z_i) \cdot h_{i-1} + z_i \cdot \hat{h}_i \\
   z_i &= \sigma(W_z s_i+U_z h_{i-1}), \\
   \hat{h}_i &= \tanh(W s_i+U(r_i \cdot h_{i-1})), \\
   r_i &= \sigma(W_r s_i + U_r h_{i-1}),
  \end{aligned}
  &&
  \begin{aligned}
   &\text{for}~~ i=1,..,m, \\ 
   &(update~~gate) \\
   &(candidate~~gate) \\
   &(reset~~gate)
  \end{aligned}
  \label{eqn:hidden state}
\end{align}
where $W_z, U_z, W, U, W_r$ and $U_r$ are weight matrices and $\sigma(\cdot)$ is the sigmoid function. The reset gate makes sure to forget information about the past that is not important given the current session. The update gate decides whether the current session contains relevant information that should be stored. The hidden state, $h_i$, is a linear interpolation between the previous hidden state and the candidate gate.

Our second way of passing input sessions through the RNN is called \emph{Cross-sessions Concat} (see Figure~\ref{fig:architecture_concat}). Here, we concatenate all sessions of a user into a single sequence $s=\lbrace a_{11},..,a_{1 n},a_{21},..,$ $a_{2 n},..,a_{m1},..,a_{m n}\rbrace$. Now the hidden state in Equation~\eqref{eqn:hidden state} is computed for every time step $(ij)$ in $s$. Whereas the cross-sessions encode only accounts for the order of sessions, the cross-sessions concat further takes into account the order of actions.

Finally, we propose another variation of session encoding, \emph{Cross-sessions Auto} (see Figure~\ref{fig:architecture_auto}), which automatically learns encodings of sessions with an autoencoder, instead of Equation~\eqref{eqn:encoding}. We train an RNN-based autoencoder with a single GRU layer that takes as input the ordered sequence of actions in a session and is evaluated on the task of recreating the input using categorical cross-entropy loss on each of the $3$ features: action section, action object and action type. Once trained, the encoder is used to encode a session into a single vector, $s_i$, that can be used as input for Equation~\eqref{eqn:hidden state}. Hence, the architecture of the autoencoder in Figure~\ref{fig:architecture_auto} is combined with the architecture in Figure~\ref{fig:architecture_encode}.

\paragraph{Cross-Entropy Loss} In all three cases, the RNN returns an output vector, $o$, of length $K$ after the last time step. Because a user can buy multiple items at the same time, we consider the learning task as multi-label classification and use the sigmoid function, $\sigma(\cdot)$, on each element of $o$ as output activation function to compute the likelihood of purchase:
\begin{equation}
    \hat{p}_k = \sigma(o_k), ~~\text{for}~~ k=1,\ldots,K.
    \label{eq:output_activation}
\end{equation}
During training, the loss function is computed by comparing $\hat{p}$ with the binarized vector of the items purchased, $p$. Due to the learning task being multi-label classification, we define the loss function as the sum of the binary cross-entropy loss over all items. The loss function is thereby different from the ranking loss used in GRU4REC and is given by:
\begin{equation}
    L = - \sum_{k=1}^K  p_k \cdot \log (\hat{p}_k) +(1-p_k) \cdot \log (1-\hat{p}_k) .
    \label{eq:loss_function}
\end{equation}

\paragraph{Censored Weibull Loss} Because users often have multiple sessions before they purchase, and at the time of prediction we do not know when they will purchase, we furthermore propose another loss function that takes the time of purchase into account. Instead of predicting the probability of each item being purchased after the last time step, we now predict the time to the next purchase (time-to-purchase) of each item after every time step in the input sequence of the user's sessions, where time is measured in days. The loss function should then compare the predicted time-to-purchase with the true time-to-purchase. However, only some of the true times are observed, since a user typically has not purchased all of the items by the end of the training period. This is what is called censored data since it is only partially observed. For that reason, we use the loss function presented in~\citet{Martinsson2017} that takes into account censored time data.

\begin{figure*}[tb]
\centering
\begin{minipage}[t]{.33\textwidth}
    \includegraphics[width=\linewidth]{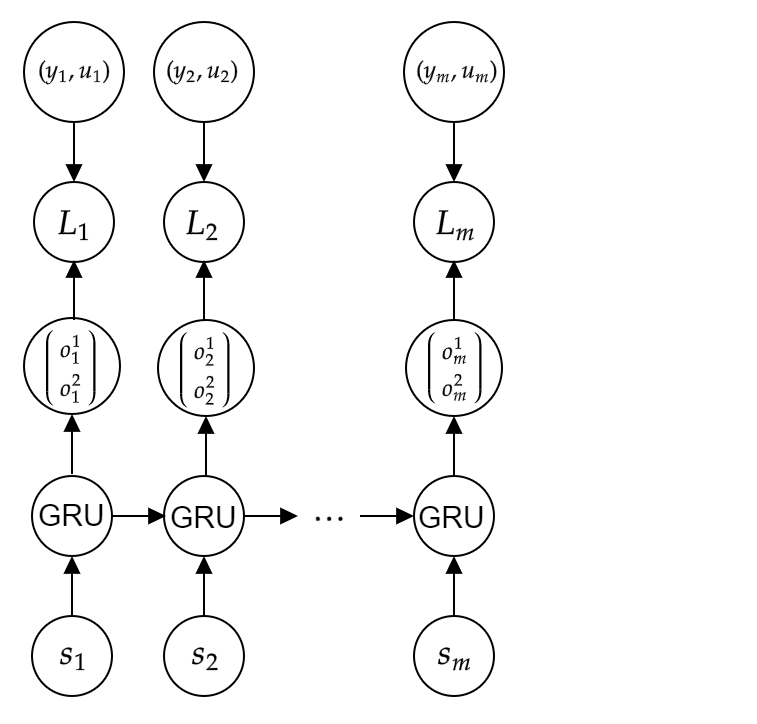}
    \caption{Architecture of Cross-sessions $~$ $~$ $~$ Encode with censored Weibull loss.}
    \label{fig:architecture_weibull_loss}
\end{minipage}%
\begin{minipage}[t]{.33\textwidth}
    \includegraphics[width=\linewidth]{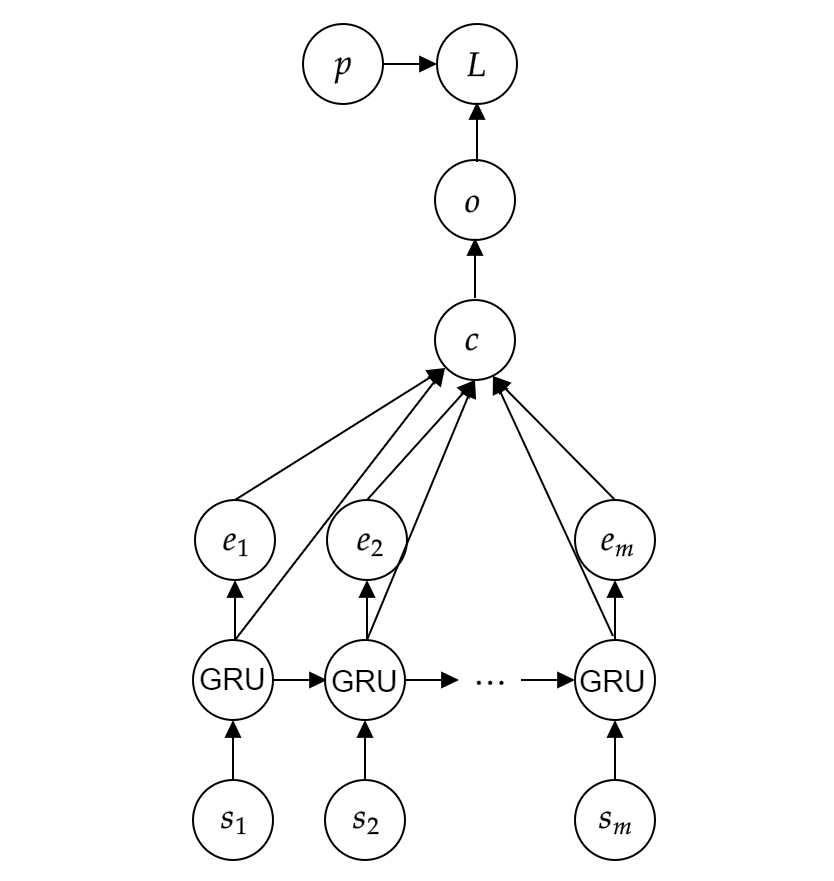}
    \caption{Architecture of Cross-sessions $~$ $~$ $~$ Encode with attention mechanism.}
    \label{fig:architecture_attention}
\end{minipage}%
\begin{minipage}[t]{.33\textwidth}
    \includegraphics[width=\linewidth]{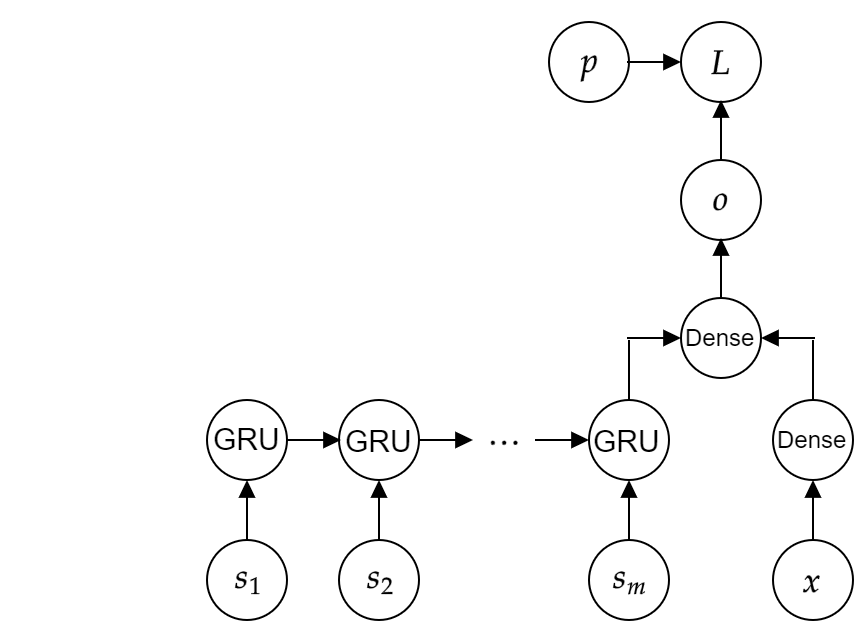}
    \caption{Architecture of a hybrid model between a cross-sessions and a demographic model. $x$ denotes an input vector representing demographic features of the user.}
    \label{fig:architecture - hybrid}
\end{minipage}
\end{figure*}

Let $y_{i,k}$ be the observed time-to-purchase of item $k$ at time step $i$. Moreover, let $u_{i,k}$ be the censoring variable that indicates whether the purchase of item $k$ has occurred after time step $i$ and before the end of the training period. We assume that $y_{i,k}$ is a realisation of the random variable $Y_{i,k}$ that follows a discrete Weibull distribution with positive parameters $\alpha_{i,k}$ and $\beta_{i,k}$ and probability mass function given by:
\begin{equation}
    P(Y_{i,k}=y_{i,k}) = \exp \Bigl( -\Bigl(\frac{y_{i,k}}{\alpha_{i,k}}\Bigr)^{\beta_{i,k}} \Bigr) - \exp \Bigl( -\Bigl(\frac{y_{i,k}+1}{\alpha_{i,k}}\Bigr)^{\beta_{i,k}} \Bigr),
    ~\text{for}~ y_{i,k} = 0,1,2,...
\end{equation}
We use the Weibull distribution as it is commonly used to model time-to-event data because it is positive, has infinite support, and contrary to other time-to-event distributions, like the exponential distribution and log-logistic distribution, the Weibull distribution has a discrete version, that can be used when time is measured in, for example, days.
The RNN returns two output vectors $o^1_i$ and $o^2_i$, both of length $K$ for every time step $i=1,...,m$, which are then transformed with an output activation function into valid parameters of the Weibull distribution (i.e., positive values). We use an exponential activation function to compute $\alpha_{i,k}$ as it has shown to give fast training time in~\citet{Chen2018} due to logarithmic effect of change in $\alpha_{i,k}$, and we use a sigmoid activation function to compute $\beta_{i,k}$ as we want slow changes when $\beta_{i,k}$ gets close to $0$:
\begin{equation}
    \begin{pmatrix} \alpha_{i,k} \\ \beta_{i,k} \end{pmatrix} = \begin{pmatrix} \exp(o^1_{i,k}) \\ \sigma(o^2_{i,k})
    \end{pmatrix}, ~\text{for}~ k=1,...,K ~\text{and}~ i=1,...,m.
\end{equation}
The loss function is given by:
\begin{equation}
    L_i = -\sum_{k=1}^K u_{i,k} \log \Bigl( P(Y_{i,k}=y_{i,k})\Bigr)+(1-u_{i,k})\log\Bigl(P(Y_{i,k}>y_{i,k})\Bigr), ~\text{for}~ i=1,...,m,
\end{equation}
where $P(Y_{i,k}>y_{i,k})$ is the right tail probability given by:
\begin{equation}
    P(Y_{i,k}>y_{i,k}) = \exp \Bigl( -\Bigl( \frac{y_{i,k}+1}{\alpha_{i,k}} \Bigr)^{\beta_{i,k}} \Bigr).
\end{equation}
Thus the loss function is given by the negative log-likelihood, where the likelihood is defined as the probability mass under the estimated parameters $\alpha_{i,k}$ and $\beta_{i,k}$ when the true time-to-purchase of item $k$ is uncensored ($u_{i,k}=1$), and the likelihood is given by the probability of the purchase to occur at some point after the end of training period when the true time-to-purchase is censored ($u_{i,k}=0$). Finally, the loss is summed up over all items. The architecture of Cross-sessions Encode with censored Weibull loss is illustrated in Figure~\ref{fig:architecture_weibull_loss}. The architecture of Cross-sessions Concat is similar, but the input is as in Figure~\ref{fig:architecture_concat} and the censored Weibull loss is computed for every action instead of every session. The architecture for Cross-sessions Auto with censored Weibull loss is the same as the one for Cross-sessions Encode, but combined with the architecture in Figure~\ref{fig:architecture_auto} to encode the input sessions.
In the recommendation phase, the time-to-purchase for each item is computed as the median of the Weibull distribution under the predicted $\alpha_{i,k}$ and $\beta_{i,k}$ parameters. The score for each item is then given by the negative time-to-purchase, such that the shorter time-to-purchase of an item the higher the score.

\paragraph{Attention Model} We propose to extend our cross-sessions models with an attention mechanism to account for the importance of different time steps in the input sequence. For instance, more weights could be given to more recent input passed to the RNN. We do this by adding an attention mechanism on the top of the GRU layer in our models that automatically learns attention weights. We use the Bahdanau attention introduced in~\citet{Bahdanau2015} to permit the decoder of the network to use the most relevant parts of the input sequence, by a weighted combination of all the hidden states, with the most relevant states being given the highest weights. 
Formally, instead of returning an output vector $o$ after the last time step in the recurrent layer, the RNN returns attention scores $e_i$ for each time step $i$. The attention scores are then normalised into attention weights, $\lambda_i$, using a softmax function:
\begin{equation}
    \lambda_i = \frac{\exp(e_i)}{\sum_{i}\exp(e_i)}.
    \label{equ:attention_weights}
\end{equation}
Subsequently, a context vector $c$ is computed as the sum of all the hidden states weighted by the attention weights:
\begin{equation}
    c = \sum_{i} \lambda_i h_i .
\end{equation}
The output vector $o$ of length $K$ is now returned from this attention layer and activated with the sigmoid function as in Equation~\ref{eq:output_activation}. The architecture of Cross-sessions Encode with attention mechanism is illutrated in Figure~\ref{fig:architecture_attention}, where the loss function is cross-entropy as in Equation~\eqref{eq:loss_function}.

\paragraph{Hybrid Model} Finally, we propose a hybrid of a cross-sessions and a demographic model where the hidden state from the cross-sessions' RNN is merged with the hidden state from a feed-forward neural network with demographic input features of the user. This concatenation is then passed through a dense layer. The architecture of Cross-sessions Encode combined with demographics is illustrated in Figure~\ref{fig:architecture - hybrid}, where the loss function is cross-entropy as in Equation~\eqref{eq:loss_function}. The combination with demographics is similar in the other cases.


\section{Dataset}

To the best of our knowledge, there is only one publicly available dataset that satisfies the criteria of our set-up~\cite{BruunEtAl2022}, specifically: (1) item scarcity; (2) target action happening outside the session; and (3) different types of actions which might not be directly associated to an item or a purchase event.
In the following, we describe such dataset (Section~\ref{subsec:dataset}), we present how to estimate the temporal threshold to deem two sessions as belonging to the same task (Section~\ref{subsec:threshold}), and we describe data pre-processing (Section~\ref{subsec:preprocessing}).

\subsection{Dataset Description}
\label{subsec:dataset}

\begin{table}[tb]
\centering
        \caption{Main properties of the dataset (*mean/std).}
        \begin{tabular}{@{}lr@{}}
            \toprule
            Users                        & 44,434     \\
            Items                        & 16         \\
            Purchase events              & 53,757     \\
            Sessions                     & 117,163    \\
            Actions                      & 1,256,156  \\ \midrule
            Purchase events per user*    & 1.21/0.51  \\
            Sessions before purchase event* & 2.18/1.68  \\
            Actions per session*         & 10.72/7.85 \\ \bottomrule
        \end{tabular}
        \label{tab:Dataset}
\end{table}

\begin{figure}[tb]
    \centering
    \includegraphics[width=0.4\textwidth]{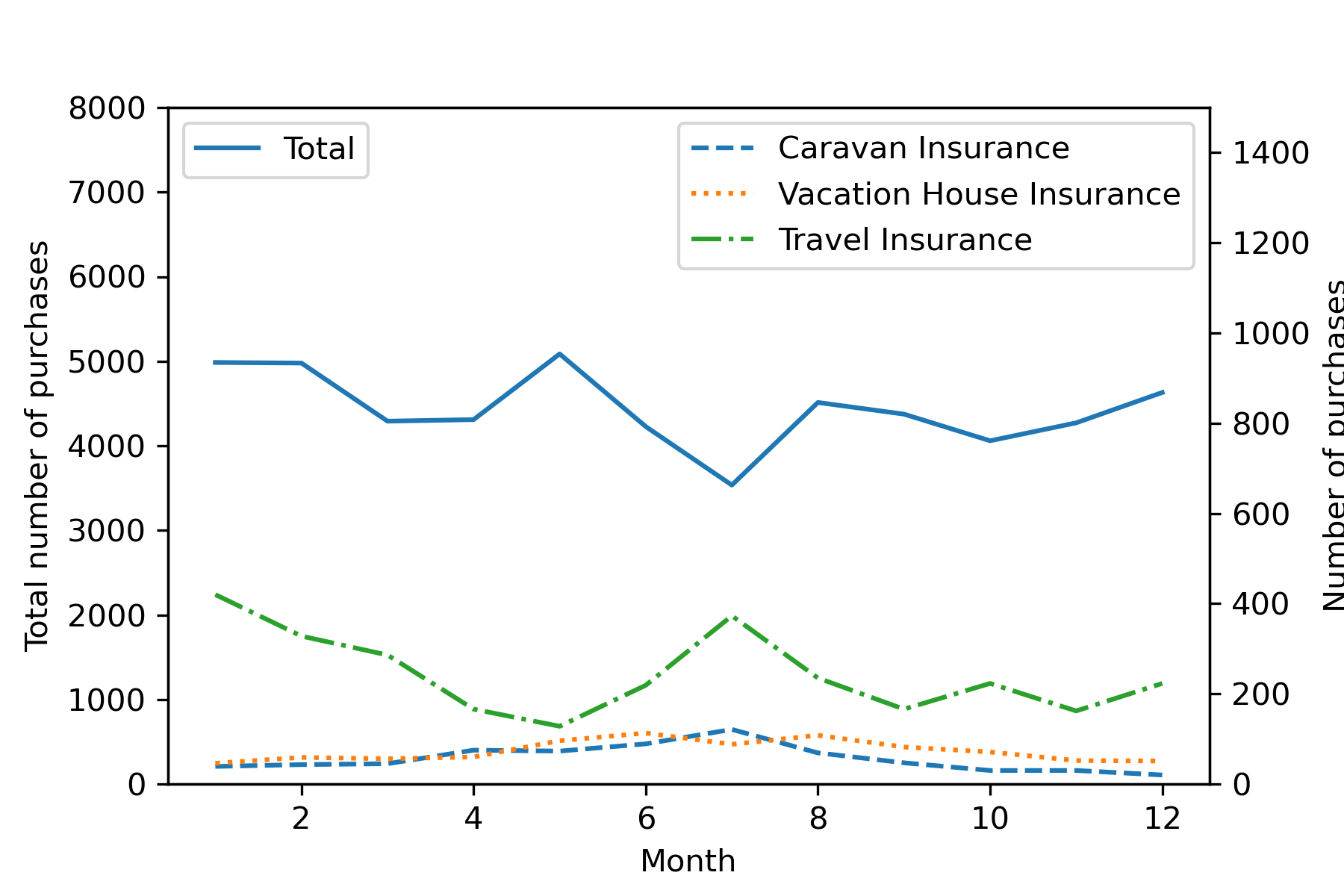}
    \caption{Number of purchases per month for all items (total) and for three selected items (caravan, vacation house and travel insurance).}
    \label{fig:season_effect}
\end{figure}

We use a publicly available dataset for the insurance domain\footnote{\url{https://github.com/simonebbruun/cross-sessions_RS/tree/main/extended}}. 
Table~\ref{tab:Dataset} reports overall statistics about the dataset.
The dataset consists of user logs on an insurance vendor website collected between October 1, 2018 and September 30, 2020.
During this period no recommender system was implemented on the insurance website that could have affected the user behavior (e.g., exposure bias or position bias~\cite{ChenEtAl2022}). 
The dataset includes interaction logs of $44$K users, who were uniquely identified either by log-in or cookies. 
There is a total of $16$ items, including insurance products or additional insurance coverages. 
An additional coverage can be bought only if the customer already has the corresponding base product. 
There are around $53$K purchase events that happened online or over the phone. 
Around $75\%$ of the users prefer to complete their purchases over the phone instead of online.
Figure~\ref{fig:season_effect} shows that there is no strong seasonal effect over the purchase frequency that the model should account for. Even caravan insurance, vacation home insurance and travel insurance have only minor seasonal trends. 

Note that this dataset differs from other publicly available datasets typically used to evaluate session-based RSs, Last.fm, Recsys Challenge (RSC) datasets, etc.
First, the total number of items is $16$, much lower than the $91$K items in Last.fm or the $29$K items in RSC15.
Second, insurance customers interact less frequently with the insurance website, indeed there are on average $1.2$ purchases per user and $2.2$ sessions before each purchase over a period of $2$ years.
Third, due to this scarcity of user interactions, the dataset is a collection of not only clicks and purchases in connection with items, but all types of actions, even if they are not directly associated with items.
Therefore, this is one of the few publicly available datasets which contains several types of actions.

\begin{table}[htb]
\centering
\caption{Actions included in the dataset with their description and frequency.}
\resizebox{\textwidth}{!}{%
\begin{tabular}{@{}lllr@{}}
\toprule
\multirow{4}{*}{Action section} & E-commerce & \begin{tabular}[c]{@{}l@{}}Includes all products and their information, description, price, etc.\\ Users can purchase products in this section.\end{tabular} & 256,319 (20.41\%) \\ \cmidrule(l){2-4} 
 & Claims reporting & Contains a form where users can report claims. & 11,188 (0.89\%) \\ \cmidrule(l){2-4} 
 & Information & Information about e.g., payment methods and contact details. & 198,147 (15.77\%) \\ \cmidrule(l){2-4} 
 & Personal account & \begin{tabular}[c]{@{}l@{}}Users can log in and see their current insurances and claims in progress,\\ change their personal information and adjust e.g., deductibles of current\\ insurances.\end{tabular} & 790,502 (62.93\%) \\ \midrule
\multirow{3}{*}{Action object} & Items & Insurance product or additional coverage. & 249,378 (19.85\%) \\ \cmidrule(l){2-4} 
 & Services & \begin{tabular}[c]{@{}l@{}}All objects that are not items, e.g., payment methods and\\ personal information.\end{tabular} & 655,574 (52.19\%) \\ \cmidrule(l){2-4} 
 & No object & \begin{tabular}[c]{@{}l@{}}Click, without any object,\\ e.g., a tab, drop down menu or front pages.\end{tabular} & 351,204 (27.96\%) \\ \midrule
\multirow{4}{*}{Action type} & Click & Click on the webpage. & 811,747 (64.62\%) \\ \cmidrule(l){2-4} 
 & Start & Start of e.g., purchase or claims report. & 388,215 (30.90\%) \\ \cmidrule(l){2-4} 
 & Act & Add product to cart, fill out claims report, etc. & 47,713 (3.80\%) \\ \cmidrule(l){2-4} 
 & Complete & Complete e.g., purchase or claims report. & 8,481 (0.68\%) \\ \bottomrule
\end{tabular}
}%
\label{tab:actions}
\end{table}

For each user-purchase pair, all sessions occurring before the purchase event are collected. 
This corresponds to around $100$K sessions. 
A session (see Section~\ref{subsec:problem}) is represented as a sequence of actions sorted by their timestamp, where each action has $3$ different components:
\begin{itemize}
    \item Action section: e-commerce, claims reporting, information and personal account;
    \item Action object: item and service:
    \item Action type: click, start, act and complete.
\end{itemize}
Table~\ref{tab:actions} provides an overview of the actions and their frequency in the dataset.
Most of the actions ($63\%$) occur in the personal account section because the users are existing customers, and most of them log in to their personal account. 
The second most frequent section is the e-commerce section ($20\%$) because items are displayed in this section and sessions are collected before purchase events.
Among the objects, the most frequent are services ($52\%$). 
The insurance website allows the user to access a number of administrative services, for example, specifying or updating the employment type, this is needed for accident insurances, or specification of annual mileage, required by car insurances, or information about the insurance coverage when moving to a new house.
Actions can happen without a specified object, in this case, they are denoted as ``no object'' ($30\%$).
This happens when a user interacts with a section, but without an object in the section, for example when a user enters the front page of a section.
Clicks are the most frequent action type ($64\%$).
This does not surprise as clicks are the primary mean of interaction with a website.
The second most frequent action is ``start'' ($31\%$), which occurs when a user starts for example a purchase or a claims report.
The action types ``act'' and ``complete'' are rather infrequent (less than $5\%$ together).
Examples of actions ``act'' are ``change'', when a user changes the employment type on the personal account page, or ``fill out'' when a user fills a form to report a claim.
The action ``complete'' occurs when a user completes a change, for example, when a user completes the change of employment or the report of a claim.

Besides user actions and purchases, the dataset includes additional features, namely demographic attributes and portfolios of users.
Demographic attributes are aggregated at a geographic area level, meaning they represent the average attribute of people living in the same area and not the exact value for every single user.
The dataset includes the following demographic attributes: age, employment, income, residence, marital status, children and education.
User portfolios include the purchase history of users within the insurance company, that is all items bought by each user.
Our cross-sessions models exploit only user interactions (sessions and purchases) and do not use the demographic attributes or the user portfolios.
However, these are needed by some state-of-the-art RS models, as SVD or the demographic model (see Section~\ref{subsec:exp_set_up}).

\subsection{Estimation of Session Threshold}
\label{subsec:threshold}

We consider a user task as a sequence of sessions that belong to the same task, that is the same user need, which eventually concludes in a purchase.
Next, we explain how to deem two sessions as belonging to the same task.
We need to estimate a temporal threshold $t$, such that given two subsequent sessions $s_1$ and $s_2$, if:
\begin{equation}
    \mathtt{start\_time}(s_2) - \mathtt{start\_time}(s_1) \leq t \ \implies \ s_1 \text{ and } s_2 \text{ belong to the same task.}
    \label{eq:threshold}
\end{equation}
That is, if the elapsed time between $s_1$ and $s_2$ is lower or equal to $t$, we will consider the two sessions as belonging to the same task.  

This problem is similar to the one of estimating session boundaries for web users.
It is straightforward to get the starting time of a user session, but it is not always possible to get the end time,
for example, whenever a user leaves a browser window open and goes ahead with other non-related tasks, for example checking emails~\cite{JansenEtAl2006,JansenEtAl2007}.
A rule of thumb for web sessions is to consider the session as concluded after $30$ minutes of inactivity.  
A similar estimate does not exist for inactivity time between sessions, therefore we estimate the threshold $t$ directly from log data.

\begin{figure}[tb]
    \centering
    \includegraphics[width=0.4\textwidth]{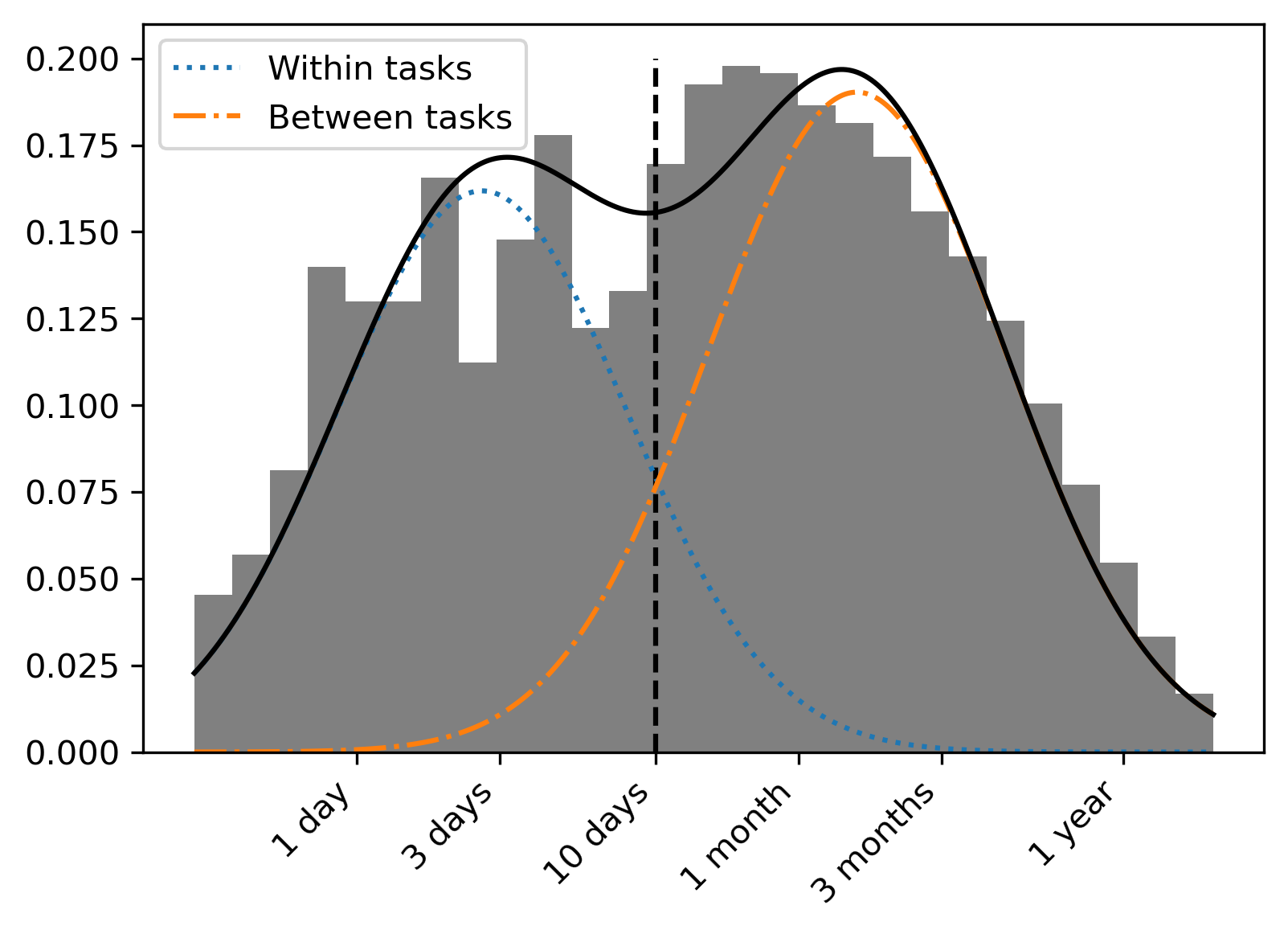}
    \caption{Histogram of logarithmically scaled inter-session times and fitted Gaussian mixture model.}
    \label{fig:inter-session times}
\end{figure}

\citet{HalfakerEtAl2015} propose a data-driven approach to estimate the inactivity time for web sessions.
We apply the same approach to estimate the task inactivity time between two consecutive sessions.
We start by computing the inter-session times between $2$ consecutive sessions from the same user. 
This is the difference between the starting time of the most recent session and the older session (see Equation~\eqref{eq:threshold}).
Figure~\ref{fig:inter-session times} shows the logarithmically scaled histogram of the inter-sessions times for all users in the dataset.
We can observe a bi-modal distribution with a valley, similar to inter-action times for web users in~\citep{HalfakerEtAl2015}.
We use Expectation Maximization (EM) to fit a two-component Gaussian mixture model.
We assume that the distribution of inter-session times is a mixture of times corresponding to: (1) sessions belonging to the same task, i.e., the blue line in Figure~\ref{fig:inter-session times}, and (2) sessions belonging to different tasks, i.e., the orange line in Figure~\ref{fig:inter-session times}.
The intersection point between the blue and orange lines represents the point where an inter-session time belongs to the two distributions with equal probability.
We use this point, which corresponds to $10$ days, as the threshold $t$.
This means that $2$ sessions belong to the same task if no more than $10$ days elapsed between their starting times.

\subsection{Dataset Pre-processing}
\label{subsec:preprocessing}

In the following, we describe all pre-processing steps.
Tables~\ref{tab:Dataset} and~\ref{tab:actions}, and Figure~\ref{fig:season_effect} are computed with the pre-processed dataset.

We removed items and actions that occur with very low frequency because they are not optimal for modeling.
We remove all items with a frequency lower than $0.1\%$, and from actions, all sections, objects, and types with a frequency lower than $0.1\%$.
Consecutive repeated actions of the same type, for example two clicks on the same object and section, are removed because they very likely represent noise due to the latency of the website or other connection issues.
Very short sessions are poorly informative, thus we remove all sessions with less than $3$ actions.
On the other side, very long sessions lead to long training times, thus we truncate all sessions in the end to have a maximum of $30$ actions (the $95$th percentile).
Based on the $10$ days threshold estimated in Section~\ref{subsec:threshold}, we discard all sessions with inter-session time greater than $10$ days.
Then, within the $10$ days rule, we keep only the $7$ most recent sessions for each user (the $95$th percentile), which allows to reduce training times. 
We keep all users, even those with a single session and purchase.

\section{Experiments}
In this section, we outline the experimental set-up (Section~\ref{subsec:exp_set_up}) as well as present and analyse the results (Section~\ref{subsec:results}).
Our source code is publicly available\footnote{\url{https://github.com/simonebbruun/cross-sessions_RS/tree/main/extended}}.

\subsection{Experimental Set-up}
\label{subsec:exp_set_up}
We start by describing the evaluation procedure, then the baseline models and finally our implementation and hyperparameter tuning.

\paragraph{Evaluation Procedure}
As a test set, we use the latest $10\%$ of purchase events with associated past sessions. The remaining $90\%$ is used for training. 
Since some users have had multiple purchase events, we remove purchase events from the training data, if their associated past sessions also appear in the test set. This resulted in $54$ out of $48.381$ purchase events in the training set being removed.

The models generate a score for how likely the user will buy each item, which is then sorted as a ranked list (i.e., the closer to the top of the ranking, the higher the estimated score of the item).
Among the two types of items, new insurance products and additional coverages, it is only possible for a user to buy an additional coverage if the user has the corresponding base insurance product. For that reason, we use a post filter to set the score to the lowest score if that is not the case, as per~\citet{Aggarwal2016}.
The list of ranked items is evaluated with Hit Rate (HR), Mean Reciprocal Rank (MRR), Precision, Recall and Mean Average Precision (MAP).
We use a cutoff threshold of $3$ because: (1) the total number of items is $16$, therefore high cut-offs (e.g., $\geq10$), will increase recall and all measures will reach high values, which will not inform on the actual quality of the RSs; (2) on the user interface the user will be recommended up to $3$ items.
Additionally, we report HR and MRR scores for all cut-offs value from $1$ to $5$.

Experimental results are supported by statistical testing. For HR we use McNemar's test \cite{Dietterich1998} and for all other measures we use one-way ANOVA \cite{kutner2005}, both with a confidence level of $0.05$, and post hoc tests to control the family-wise error rate due to multiple comparisons.

\paragraph{Baselines} We compare our models against the following state-of-the-art baselines:
\begin{itemize}
    \item \textbf{Random} recommends random items to the user.
    \item \textbf{Popular} recommends the items with the largest number of purchases across users.
    \item \textbf{SVD} is a method that factorizes the user-item matrix by singular value decomposition~\cite{CremonesiEtAl2010}. The portfolio data forms the user-item matrix, where a user-item entry is $1$ if the user has bought the item and $0$ otherwise. In the insurance domain it is likely for a user to buy the same item multiple times (e.g., a second car insurance), but matrix factorization cannot make repeated recommendations. Therefore, we add repeated items as new items (columns) to the matrix.
    \item \textbf{Demographic} is a classification model, as per~\citet{Qazi2017, Qazi2020}, that uses user demographics and their portfolios as input features. The portfolios are represented with a feature for each item counting how many of the items the user has already bought. Demographic features and portfolio features are concatenated. We use a feed forward neural network to make a fair comparison with the neural session-based approaches.
    \item \textbf{GRU4REC} is a neural session-based model~\cite{HidasiEtAl2016,HidasiAndKaratzoglou2018} for single user sessions. As input, we use the last session of a user, consisting of the 3-tuple user actions described in Section \ref{subsec:approach}. For every time step, the model outputs the likelihood for each action to be the one the user interacts with next. Recommendations are based on the output after the final time step.
    \item \textbf{GRU4REC Concat} is the same as GRU4REC, but all recent sessions of a user are concatenated into a single session.
    \item \textbf{SKNN\_E} is a session-based nearest neighbour model as the one presented in~\citet{JannachAndLudewig2017} with the extension suggested in~\citet{Latifi2021}. The nearest neighbours are determined based on the set of actions in all recent sessions of each user. A user's set of actions is a vector computed with a maximum pooling operation over the actions generated by the user. We then adapt this baseline to our task, so the recommendations are based on the items purchased by the neighbours of the target user rather than the items interacted with in the ongoing session.
    \item \textbf{SKNN\_EB} is the same as SKNN\_E, but with a further extension suggested in~\citet{Latifi2021}: scores of items previously interacted with are boosted with a factor. The factor is tuned as a hyperparameter.
\end{itemize}
Note that the SVD and the demographic model make use of portfolio data, that is users' past purchases, while this is not the case for all cross-sessions models, GRU4REC and SKNN. GRU4REC is included as it has shown the best performance under identical conditions on various datasets among all the neural models compared in~\citet{LudewigEtAl2021} and in~\citet{Latifi2021}.

We tried the sequential extension to SKNN, Vector Multiplication SKNN, which is presented in~\citet{LudewigAndJannach2018}, but did not obtain better results than the original one. Sequence and Time Aware Neighborhood~\cite{GargEtAL2019} is not included as a baseline since it was not possible to adapt it to the task under consideration (for the reasons discussed in Section~\ref{subsec:related_session}).

\paragraph{Implementation \& Hyperparameters}

All implementation is in \texttt{Python} \texttt{3.7.4} and \texttt{Ten\-sor\-Flow} \texttt{2.6.0}\footnote{We used Tensorflow's implementation of padding and masking to deal with variable length input in the RNNs.}. We used \texttt{Adam} as the optimizer with TensorFlow's default settings for the learning rate, exponential decay rates and the epsilon parameter. Early stopping was used to choose the number of epochs based on the minimum loss on the validation set (explained below). We used two-layer networks\footnote{In all models the second layer is a dense layer with ReLU activation function.} with dropout regularization on the first hidden layer.

We extracted a validation set from the training set in the same way as we extracted the test set from the whole dataset, so the validation set includes the latest $10\%$ of purchases with associated sessions and the remaining is used for training. 
We tuned the hyperparameters of each neural model (batch size, number of units and dropout rate) on the validation set using grid search. We tested powers of $2$ for the batch size and number of units ranging from $16$ to $512$. For the dropout rate, we tested values in $[0.1,0.5]$ with step size $0.1$.
The final hyperparameters used are reported in Table~\ref{tab:hyperparameters}.

For GRU4REC and GRU4REC Concat we tried three different loss functions: cross-entropy, BPR \cite{Rendle2009} and TOP1 \cite{HidasiEtAl2016}. Cross-entropy was finally chosen for both models, as it performed best on the validation set. For the non-neural models, the optimal number of latent factors for the SVD model was $1$, the optimal number of neighbours for both SKNN\_E and SKNN\_EB was $30$, and the optimal boost factor for SKNN\_EB was $0.5$. Neural models were trained on Nvidia GeForce MX250 equipped with 2GB of GPU memory. The maximum training time was ~$6$ hours.

\begin{table}[tb]
\caption{Hyperparameters (*autoencoder/RNN)}
\begin{tabular}{@{}lrrr@{}}
\toprule
\multicolumn{1}{c}{Model} & \multicolumn{1}{c}{Batch size} & \multicolumn{1}{c}{Units} & \multicolumn{1}{c}{Dropout} \\ \midrule
Demographic & 32 & 32 & 0.3 \\
GRU4REC & 32 & 256 & 0.2 \\
GRU4REC Concat & 32 & 256 & 0.2 \\
Cross-sessions Encode & 32 & 64 & 0.3 \\
Cross-sessions Concat & 128 & 64 & 0.3 \\
Cross-sessions Auto* & 128/32 & 512/64 & -/0.4 \\
Cross-sessions Encode with Weibull Loss & 16 & 64 & 0.3 \\
Cross-sessions Concat with Weibull Loss & 128 & 128 & 0.3 \\
Cross-sessions Auto with Weibull Loss* & 128/64 & 512/128 & -/0.4 \\
Cross-sessions Encode with Attention & 32 & 64 & 0.3 \\
Cross-sessions Concat with Attention & 128 & 64 & 0.3 \\
Cross-sessions Auto with Attention* & 128/32 & 512/64 & -/0.4 \\ \bottomrule
\end{tabular}
\label{tab:hyperparameters}
\end{table}

\subsection{Experimental Results}
\label{subsec:results}

First, we compare our cross-sessions models against state-of-the-art baselines.
Then we evaluate the best cross-sessions models combined with demographics.
We further break down the performance of our models to understand the impact of the time of prediction, the number of sessions and actions, the order and recency of sessions as well as the size of the session threshold. Lastly, we conduct an ablation study to show how different actions (sections, objects and types) affect the performance of our models and a fairness analysis to show how the models perform on customers with different age, gender and income level.

\begin{table}[tb]
\caption{Performance results. All results marked with * are significantly different from Cross-sessions Encode with Attention.
The best score for each measure is in bold. Percentages in brackets denote the difference between our models and the strongest baseline (SKNN\_EB).}
\resizebox{\textwidth}{!}{%
\begin{tabular}{@{}lrrrrr@{}}
\toprule
\multicolumn{1}{c}{Model} & \multicolumn{1}{c}{HR@3} & \multicolumn{1}{c}{Precision@3} & \multicolumn{1}{c}{Recall@3} & \multicolumn{1}{c}{MRR@3} & \multicolumn{1}{c}{MAP@3} \\ \midrule
Random & 0.3235* & 0.1114* & 0.2940* & 0.1910* & 0.1839* \\
Popular & 0.6217* & 0.2145* & 0.5855* & 0.4764* & 0.4540* \\
SVD & 0.6646* & 0.2372* & 0.6327* & 0.4997* & 0.4829* \\
Demographic & 0.7392* & 0.2649* & 0.7095* & 0.5620* & 0.5446* \\
GRU4REC & 0.6479* & 0.2313* & 0.6208* & 0.5443* & 0.5264* \\
GRU4REC Concat & 0.6616* & 0.2365* & 0.6362* & 0.5620* & 0.5453* \\
SKNN\_E & 0.8106* & 0.2914* & 0.7848* & 0.6740* & 0.6567* \\
SKNN\_EB & 0.8132* & 0.2922* & 0.7872* & 0.6785* & 0.6610* \\ \midrule
Cross-sessions Encode & 0.8380 (3.04\%) & \textbf{0.3030 (3.67\%)} & \textbf{0.8145 (3.46\%)} & 0.7093 (4.53\%) & 0.6923 (4.73\%) \\
Cross-sessions Concat & 0.8265 (1.62\%) & 0.2984 (2.12\%) & 0.8019 (1.87\%) & 0.7051 (3.92\%) & 0.6876 (4.02\%) \\
Cross-sessions Auto & 0.8356 (2.74\%) & 0.3024 (3.48\%) & 0.8128 (3.24\%) & 0.7085 (4.41\%) & 0.692 (4.69\%) \\ \midrule
Cross-sessions Encode with Weibull Loss & 0.8378 (3.02\%) & 0.3013 (3.10\%) & 0.8120 (3.15\%) & 0.7039 (3.74\%) & 0.6852 (3.66\%) \\
Cross-sessions Concat with Weibull Loss & 0.8289 (1.92\%) & 0.2994 (2.44\%) & 0.8048 (2.23\%) & 0.7017 (3.41\%) & 0.6843 (3.53\%) \\
Cross-sessions Auto with Weibull Loss & 0.8352 (2.70\%) & 0.3031 (3.71\%) & 0.8124 (3.20\%) & 0.7076 (4.28\%) & 0.6909 (4.52\%) \\ \midrule
Cross-sessions Encode with Attention & \textbf{0.8385 (3.11\%)} & 0.3026 (3.56\%) & 0.8141 (3.41\%) & \textbf{0.7118 (4.90\%)} & \textbf{0.6941 (5.00\%)} \\
Cross-sessions Concat with Attention & 0.8317 (2.26\%) & 0.3016 (3.23\%) & 0.8090 (2.76\%) & 0.7091 (4.50\%) & 0.6928 (4.81\%) \\
Cross-sessions Auto with Attention & 0.8324 (2.36\%) & 0.3012 (3.08\%) & 0.8095 (2.83\%) & 0.7103 (4.69\%) & 0.6932 (4.87\%) \\ \bottomrule
\end{tabular}
}%
\label{tab:results}
\end{table}

\begin{table*}[tb]
\caption{The versions of our cross-sessions encode models enhanced with demographic data. The rest of the notation is as in Table~\ref{tab:results} (i.e., percentages in brackets denote the difference from the strongest baseline).}
\resizebox{\textwidth}{!}{%
\begin{tabular}{lrrrrr}
\hline
\multicolumn{1}{c}{Model} & \multicolumn{1}{c}{HR@3} & \multicolumn{1}{c}{Precision@3} & \multicolumn{1}{c}{Recall@3} & \multicolumn{1}{c}{MRR@3} & \multicolumn{1}{c}{MAP@3} \\ \hline
Cross-sessions Encode & \textbf{0.8542* (5.03\%)} & \textbf{0.3103* (6.18\%)} & \textbf{0.8313* (5.60\%)} & 0.7268* (7.11\%) & 0.7099* (7.41\%) \\
Cross-sessions Encode with Weibull Loss & 0.8529* (4.87\%) & 0.3100* (6.09\%) & 0.8310* (5.55\%) & 0.7219 (6.39\%) & 0.7055 (6.74\%) \\
Cross-sessions Encode with Attention & 0.8514* (4.69\%) & 0.3093* (5.83\%) & 0.8284* (5.23\%) & \textbf{0.7301* (7.59\%)} & \textbf{0.7134* (7.93\%)} \\ \hline
\end{tabular}
}%
\label{tab:results2}
\end{table*}

\begin{figure}[tb]
    \centering
    \begin{subfigure}{0.4\textwidth}
        \centering
        \includegraphics[width=\columnwidth]{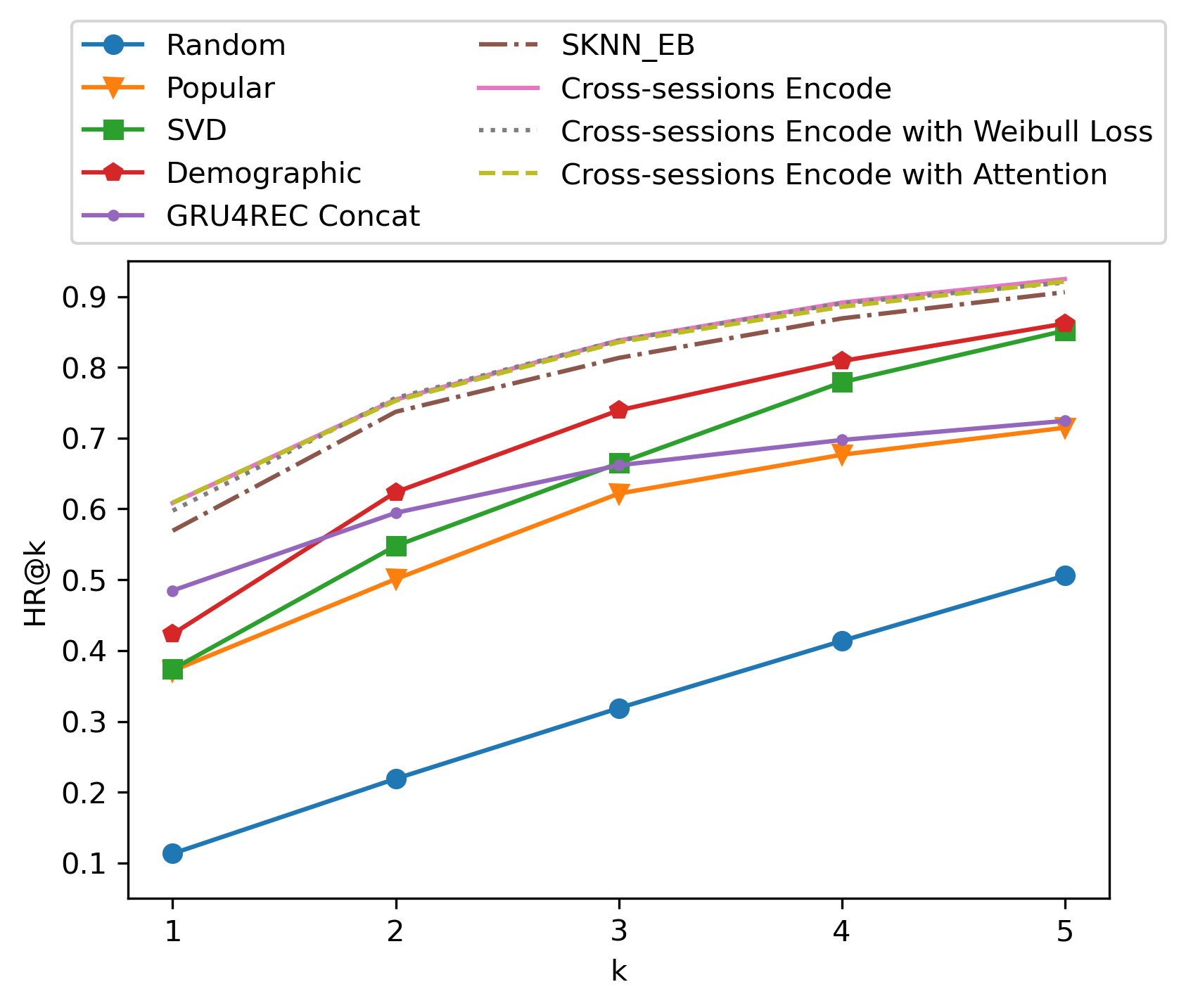}
    \end{subfigure}
    \begin{subfigure}{0.4\textwidth}
        \centering
        \includegraphics[width=\columnwidth]{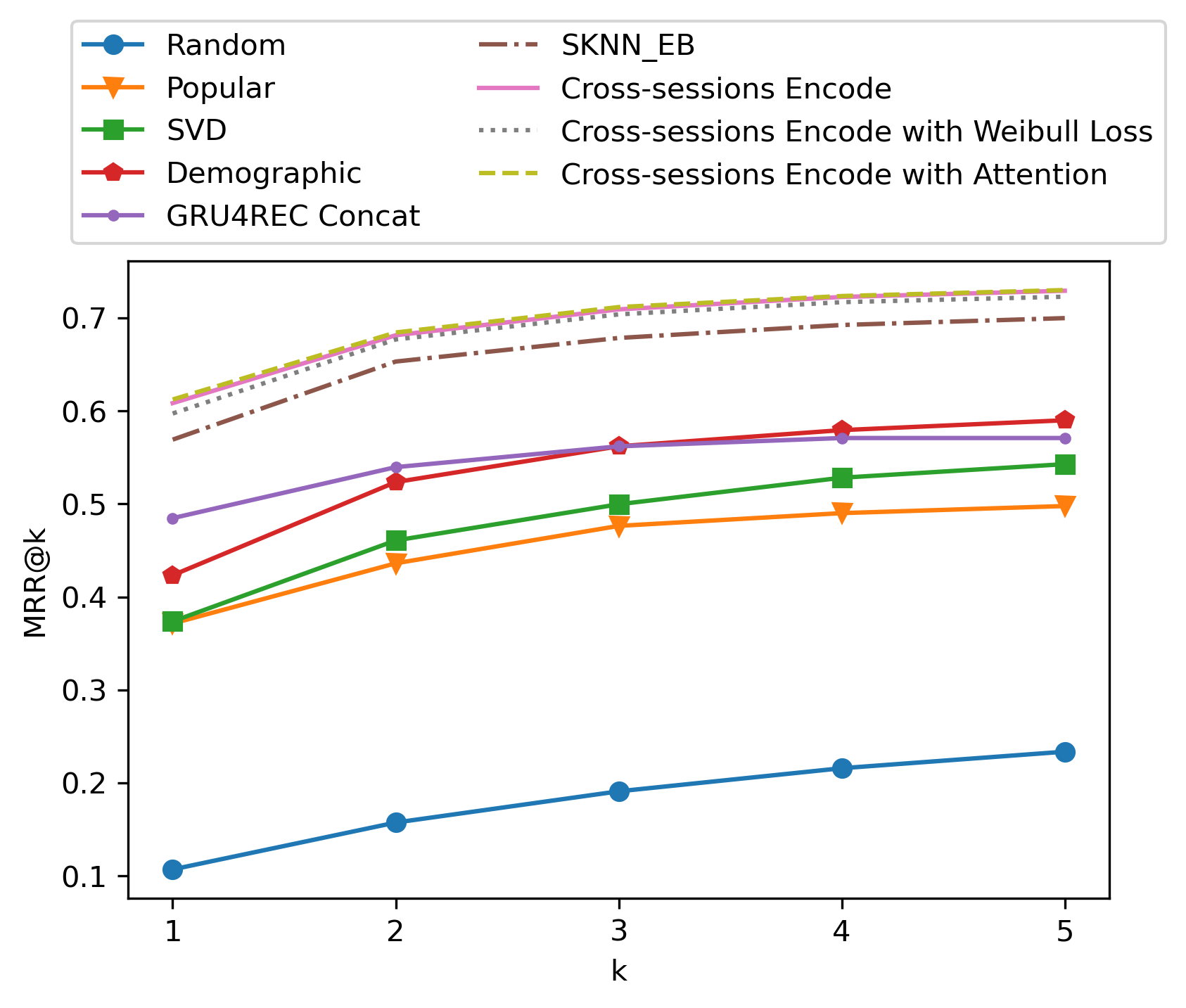}
    \end{subfigure}
    \caption{HR@k and MRR@k for varying choices of the cutoff threshold $k$.}
    \label{fig:varying cutoffs}
\end{figure}

\begin{figure}[tb]
    \centering
    \begin{subfigure}{0.4\textwidth}
        \centering
        \includegraphics[width=\columnwidth]{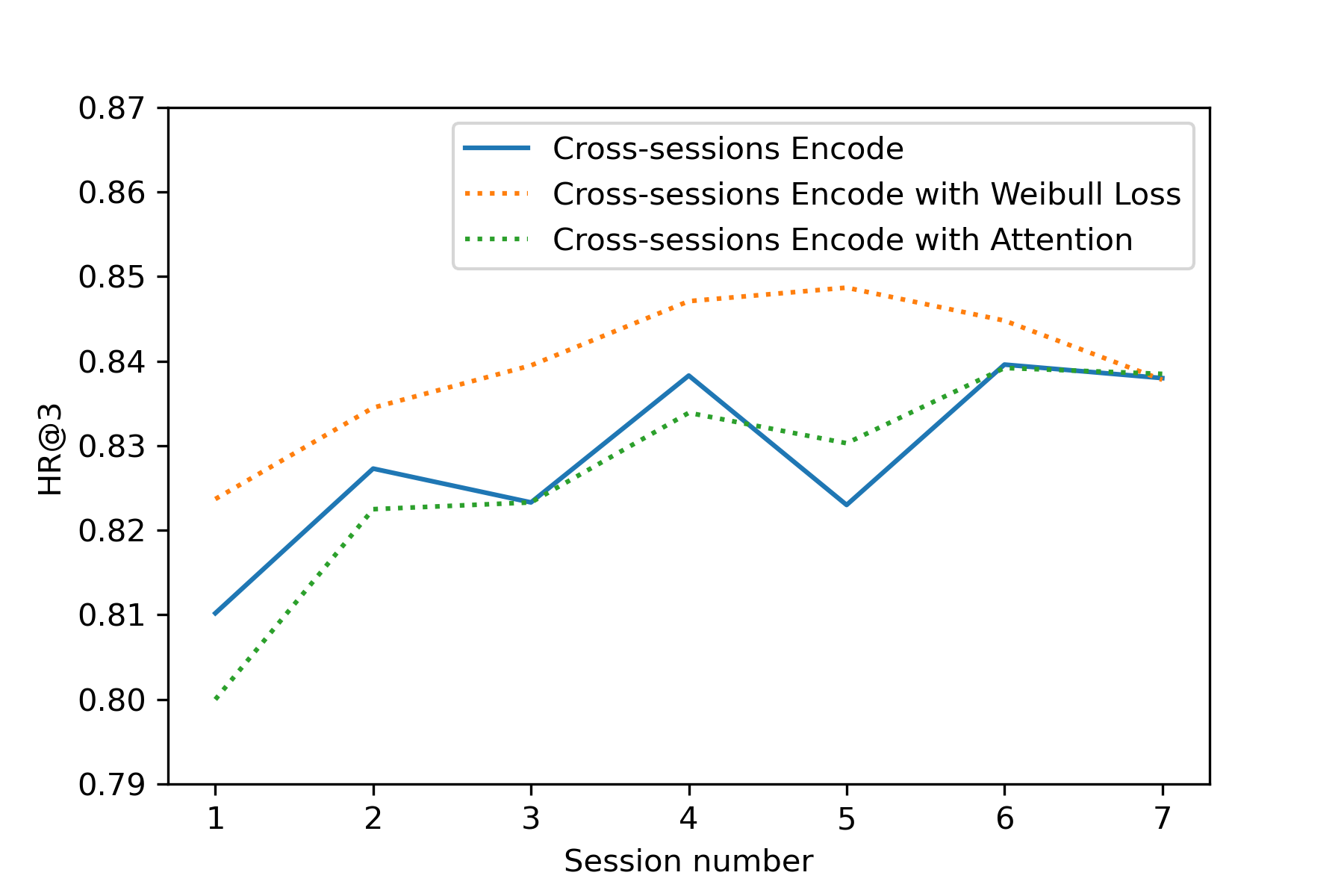}
    \end{subfigure}
    \begin{subfigure}{0.4\textwidth}
        \centering
        \includegraphics[width=\columnwidth]{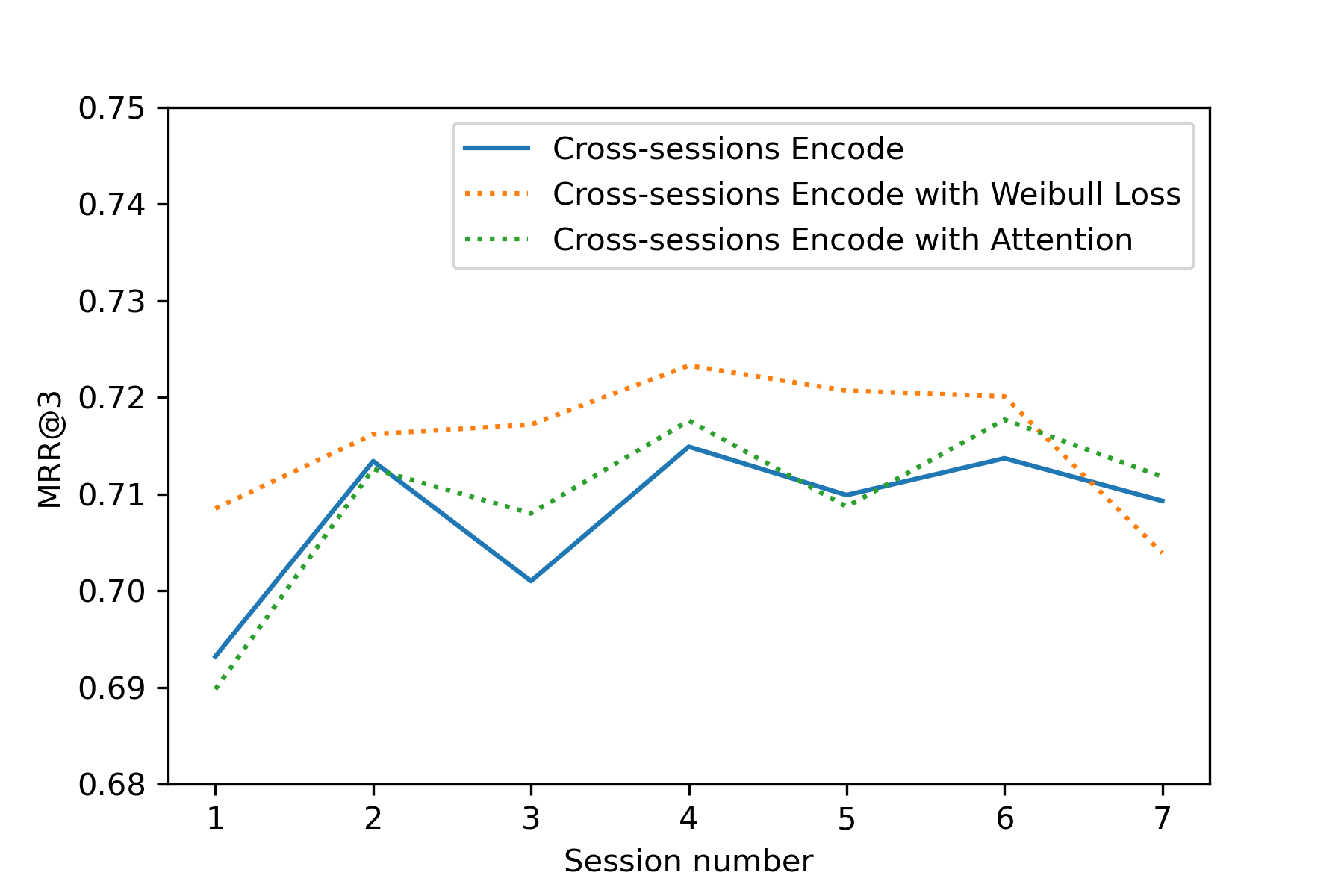}
    \end{subfigure}
    \caption{HR@$3$ and MRR@$3$ for each step in the users' sequences of sessions.}
    \label{fig:weibull}
\end{figure}

\paragraph{Performance Analysis}
Table~\ref{tab:results} presents a comparison of our cross-sessions models against the baselines. 
Unlike domains like retail and video services~\cite{HidasiEtAl2016,DacremaEtAl2021}, the simple popular model is quite a strong baseline, because of the small number of different items in the insurance dataset and the role of the post filter to make sure not to recommend items that the user cannot buy.

As seen in prior work on insurance RSs~\cite{Qazi2017}, we also see a significant improvement in using a demographic RS compared to the traditional matrix factorisation method, SVD. This is likely due to the sparse feedback on items in the insurance domain (see Table~\ref{tab:Dataset}) and the fact that users' demographic characteristics are good signals for learning insurance recommendations.

The session-based methods, SKNN\_E, SKNN\_EB and cross-sessions, significantly outperform the non-session-based methods, while this is not the case for GRU4REC and GRU4REC Concat. This shows that users' sessions are stronger signals for insurance recommendations than long-term preferences and demographic characteristics, but recommending the item that the user is most likely to interact with next on the website is not appropriate for insurance recommendations.
Moreover, all the cross-sessions models significantly outperform SKNN\_E and SKNN\_EB suggesting that an RNN is better at modeling relationships between the user actions that lead to the purchase of specific items.

The results suggest that encoding of sessions is better than the trivial concatenation of sessions indicating that dependencies across sessions are important.
The results further suggest that the encoding of sessions with a maximum pooling operation is better than an autoencoding of sessions. This is most likely because the order of actions (which the autoencoder takes into account) adds more noise to the model than signal, or the autoencoder needs a larger amount of training data in order to effectively learn to encode sessions.
The best results of our cross-sessions models trained with censored Weibull loss are obtained with Cross-sessions Encode and Cross-sessions Auto, but the models do not improve over Cross-sessions Encode trained with cross-entropy loss. Below, we further examine how the censored Weibull loss affects the performance at different times of recommendation.
The performance of the cross-sessions models enhanced with attention mechanism slightly improves over the models without attention. The degree of improvement is similar across the three variations of encodings. Below, we further explore the attention mechanism by extracting the learned attention weights.
Overall, encoding of sessions with a maximum pooling operation performed best. Hence in the rest of the analysis, we focus on Cross-sessions Encode trained with and without Weibull loss and attention mechanism.

Figure~\ref{fig:varying cutoffs} shows HR and MRR at varying cutoffs $k$ from $1$ to $5$.
We observe similar results for recall, precision and MAP.
The results are consistent over varying choices of $k$, with the exception of GRU4REC Concat which is better than SVD and Demographic for smaller cutoff thresholds (1-2), but not for larger. Across all choices of $k$ there is a clear gap between the cross-sessions models and the others.
The general trend for both measures is that they tend to increase as the cut-off $k$ increases. This is expected to happen since it is more likely to include the purchased items when $k$ increases.

The results of our cross-sessions models combined with demographics are shown in Table~\ref{tab:results2}.
This hybrid approach yields better performance than the individual models for all evaluation measures.
This indicates that the two types of information, sessions and demographic, capture different aspects of the problem.
The best results are obtained with Cross-sessions Encode and Cross-sessions Encode with Attention and both models are significantly different from the best model without demographic data.

Overall, the results show that user sessions are stronger signals for insurance recommendations than insurance portfolios and users' demographics, but recommending the item that the user is most likely to interact with next on the website is not appropriate for insurance recommendations. Moreover, an RNN is better at modeling the relationship between user actions and purchases than an SKNN, and the results suggest that encoding of sessions is better than the trivial concatenation of sessions. Finally, the results show that the two types of information, user sessions and demographics, capture different aspects of the problem.

\paragraph{Analysis of the Time of Recommendation.}
Until now, we have evaluated all models after the final step in the users' sequences of sessions, when most information is available. However, when executing the models in a real-life scenario with online users, we do not know when each user will buy and we want to make recommendations already from their first session. In Figure~\ref{fig:weibull} we compare our cross-sessions models, where we evaluate them for each step in the sequences of users' sessions. We see that the model trained with Weibull loss performs better than the models trained with cross-entropy loss for earlier steps. It is most likely because the censored Weibull loss takes into account the time of purchase. Hence, at the final step in the users' sequences of sessions, the two types of loss function result in similar performance, but the censored Weibull loss gives better results for earlier steps, which is desirable in a real-life scenario. We observe similar results for precision, recall and MAP.

\paragraph{Analysis of Number of Sessions and Actions}
\begin{figure}[tb]
    \centering
    \begin{subfigure}{0.4\textwidth}
        \centering
        \includegraphics[width=\columnwidth]{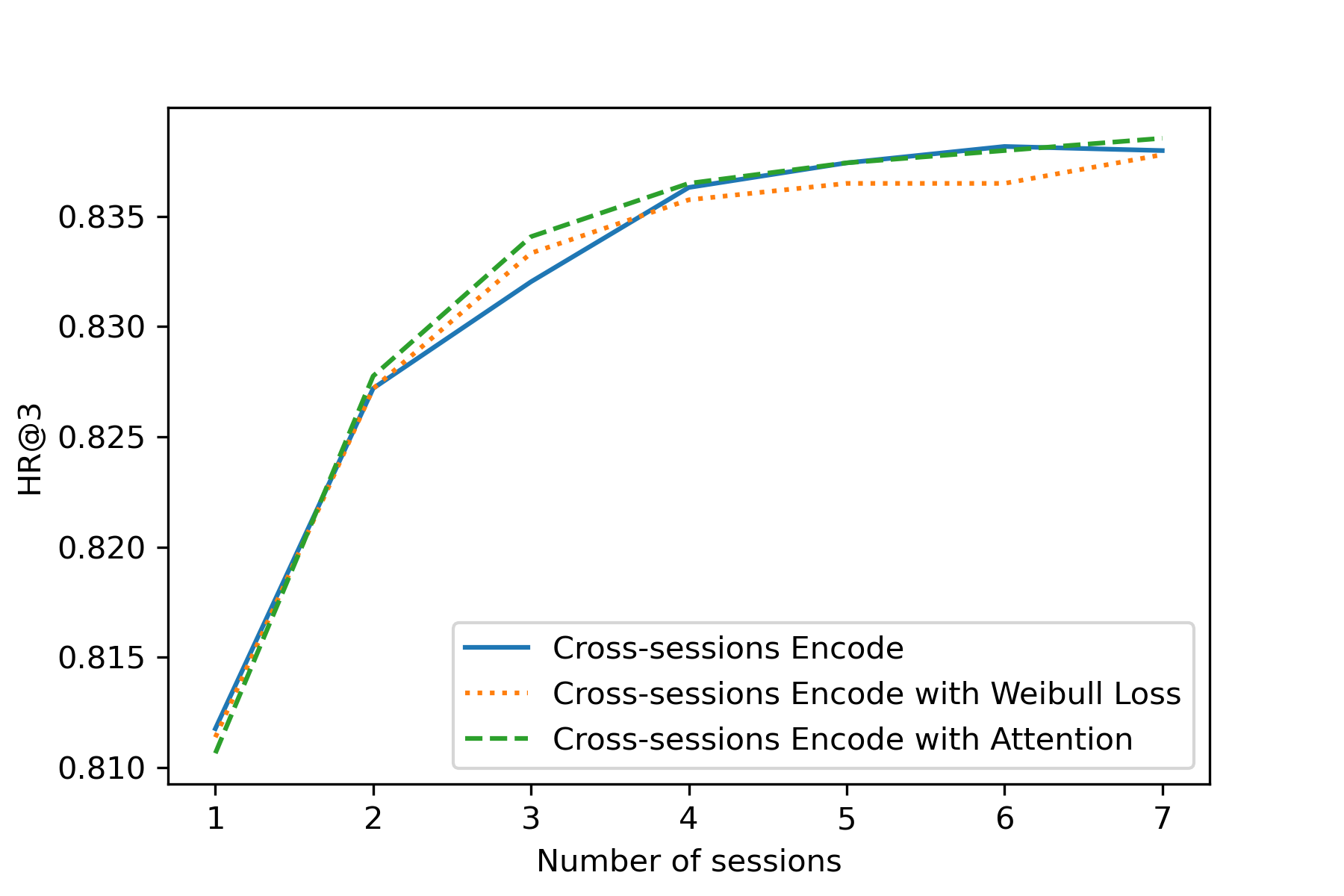}
    \end{subfigure}
    \begin{subfigure}{0.4\textwidth}
        \centering
        \includegraphics[width=\columnwidth]{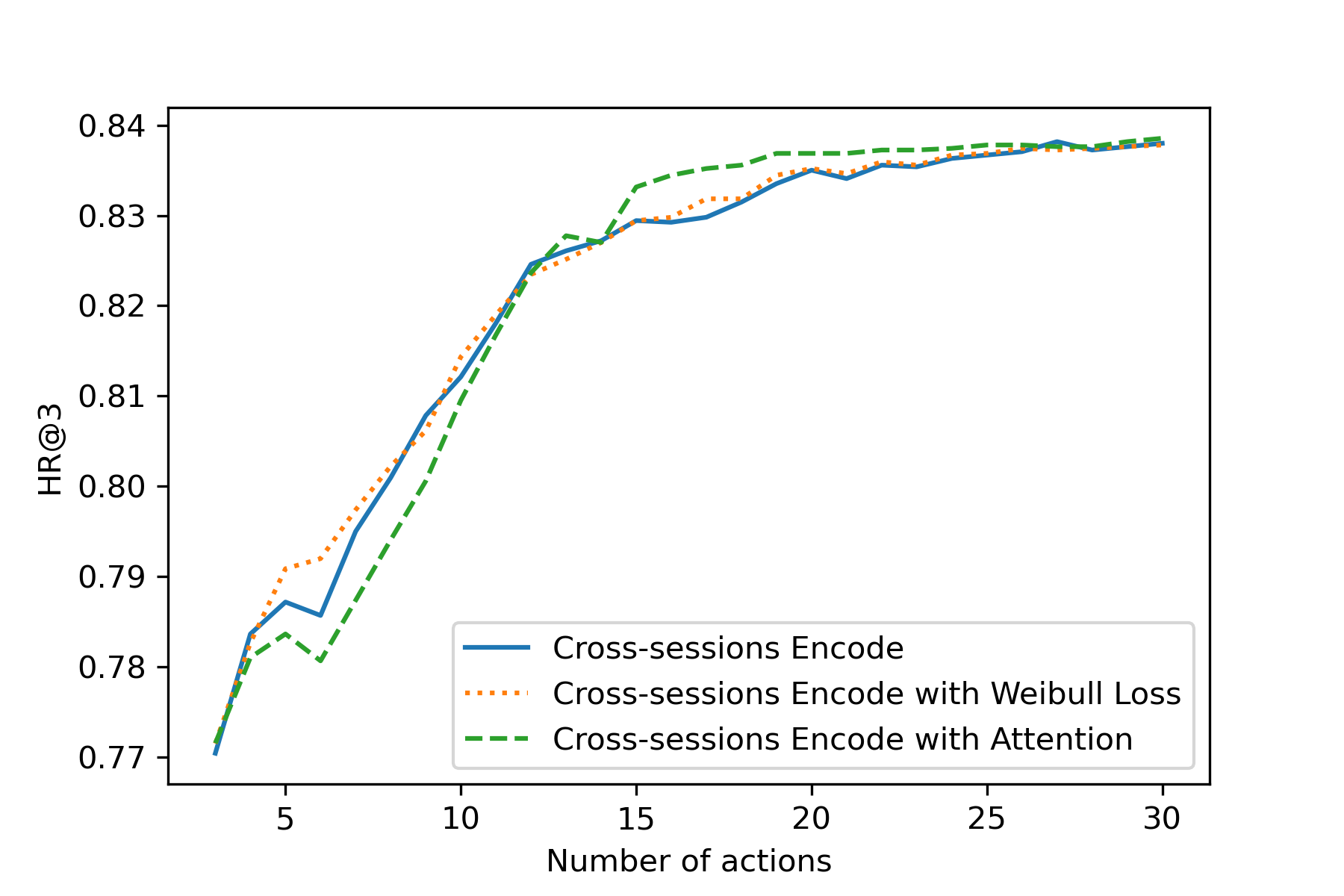}
    \end{subfigure}
    \caption{HR@3 for different number of sessions and actions.}
    \label{fig:num_sessions_actions}
\end{figure}

Next, we analyse how the number of sessions affects our cross-sessions models.
We break down the performance of our models based on the number of sessions, starting with only the most recent session, up to including all the available sessions (the maximum number of sessions per user is $7$, see Sections~\ref{subsec:threshold} and~\ref{subsec:preprocessing}).
Figure~\ref{fig:num_sessions_actions} (left) shows HR@$3$ computed for our cross-sessions models for varying numbers of user sessions.
In general, there is an increasing trend in performance with the number of recent sessions, emphasising the additional contribution brought by all recent sessions of each user rather than just the last one.
We observe similar results for MRR, recall, precision and MAP.

We do the same analysis with the number of actions per session. Figure~\ref{fig:num_sessions_actions} (right) shows that HR@$3$ generally increases with the number of actions.
The growth is particularly steep up to about $10$ actions per session after which it flattens out.
We observe similar results for MRR, recall, precision and MAP.

\paragraph{Analysis of Session Order}
We analyse the importance of session order by randomly shuffling the order of sessions, then retraining the models. We shuffle the order in both training, validation and test set, and perform the experiment $5$ times to account for randomness. The mean performance is presented in Table~\ref{tab:shuffled_session_order}. Across all our models and evaluation measures, performance drops when shuffling the session order, but the decrease is limited to less than $1.5\%$. The results indicate that the superiority of the cross-sessions models is not due to sequential dependencies, rather they are simply better at capturing the relationships between user actions and the purchase of specific items.

\begin{table*}[tb]
\caption{Study of session order. Relative change in parentheses.}
\resizebox{\textwidth}{!}{%
\begin{tabular}{@{}llrrrrr@{}}
\toprule
\multicolumn{2}{c}{Model} & \multicolumn{1}{c}{HR@3} & \multicolumn{1}{c}{Precision@3} & \multicolumn{1}{c}{Recall@3} & \multicolumn{1}{c}{MRR@3} & \multicolumn{1}{c}{MAP@3} \\ \midrule
\multirow{2}{*}{Cross-sessions Encode} & original session order & 0.8380 & 0.3030 & 0.8145 & 0.7093 & 0.6923 \\
 & shuffled session order & 0.8345 (-0.41\%) & 0.3008 (-0.7\%) & 0.8096 (-0.60\%) & 0.7058 (-0.49\%) & 0.688 (-0.61\%) \\ \midrule
\multirow{2}{*}{\begin{tabular}[c]{@{}l@{}}Cross-sessions Encode\\ with Weibull Loss\end{tabular}} & original session order & 0.8378 & 0.3013 & 0.8120 & 0.7039 & 0.6852 \\
 & shuffled session order & 0.8299 (-0.95\%) & 0.2992 (-0.68\%) & 0.8053 (-0.82\%) & 0.6991 (-0.68\%) & 0.6813 (-0.56\%) \\ \midrule
\multirow{2}{*}{\begin{tabular}[c]{@{}l@{}}Cross-sessions Encode\\ with Attention\end{tabular}} & original session order & 0.8385 & 0.3026 & 0.8141 & 0.7118 & 0.6941 \\
 & shuffled session order & 0.835 (-0.42\%) & 0.3012 (-0.46\%) & 0.8105 (-0.44\%) & 0.7038 (-1.12\%) & 0.6863 (-1.12\%) \\ \bottomrule
\end{tabular}
}%
\label{tab:shuffled_session_order}
\end{table*}

\paragraph{Analysis of Input Recency}
We analyse the importance of different time steps in the input sequence to our cross-sessions models, specifically, if more weights should be given to more recent input passed to the RNNs. We do this by extracting the attention weights, $\lambda_i$ (see Equation~\eqref{equ:attention_weights}), from the cross-sessions encode model with attention mechanism. The weights are presented in Table~\ref{tab:attentions}. The model learns an attention weight for every step in the sequence of the user's past sessions. Since not all users have had $7$ sessions (the maximum number of sessions), we present the attention weights averaged over users with the same number of sessions.
For the users with only one session, the average attention weight is simply $1$. For users with more than one session, we observe that the last session is assigned the greatest weight, while the previous sessions are weighted almost equally. This shows that the importance of input sessions does not decay linearly with the recency of the sessions, rather the last session is most important and the rest of the sessions are equally important.

\begin{table}[tb]
\caption{Attention weights averaged over users with the same number of sessions. Columns represent the number in the sequence of sessions and rows represent subsets of users with the same number of sessions.}
\begin{tabular}{@{}crrrrrrr@{}}
\toprule
\multirow{2}{*}{\begin{tabular}[c]{@{}c@{}}Total number\\ of sessions\end{tabular}} & \multicolumn{7}{c}{Session number} \\
 & \multicolumn{1}{c}{1} & \multicolumn{1}{c}{2} & \multicolumn{1}{c}{3} & \multicolumn{1}{c}{4} & \multicolumn{1}{c}{5} & \multicolumn{1}{c}{6} & \multicolumn{1}{c}{7} \\ \midrule
1 & 0 & 0 & 0 & 0 & 0 & 0 & 1 \\
2 & 0 & 0 & 0 & 0 & 0 & 0.3059 & 0.6941 \\
3 & 0 & 0 & 0 & 0 & 0.2326 & 0.2201 & 0.5473 \\
4 & 0 & 0 & 0 & 0.1942 & 0.1735 & 0.1824 & 0.4498 \\
5 & 0 & 0 & 0.1353 & 0.1509 & 0.1418 & 0.1537 & 0.4183 \\
6 & 0 & 0.1453 & 0.1040 & 0.1169 & 0.1148 & 0.1350 & 0.3840 \\
7 & 0.1243 & 0.1015 & 0.0867 & 0.1039 & 0.1112 & 0.1273 & 0.3451 \\ \bottomrule
\end{tabular}
\label{tab:attentions}
\end{table}

\begin{figure}[tb]
    \centering
    \begin{subfigure}{0.4\textwidth}
        \centering
        \includegraphics[width=\columnwidth]{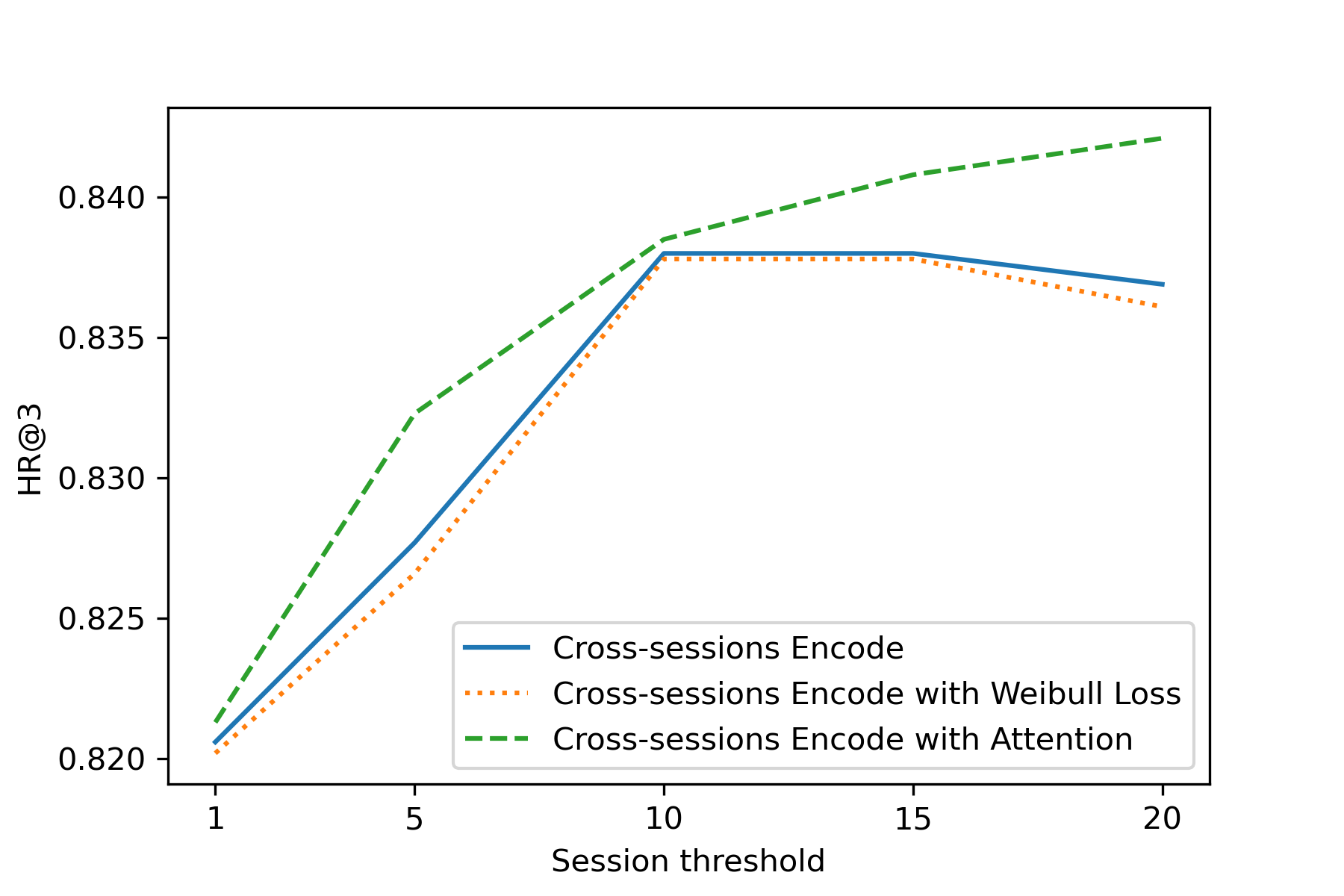}
    \end{subfigure}
    \begin{subfigure}{0.4\textwidth}
        \centering
        \includegraphics[width=\columnwidth]{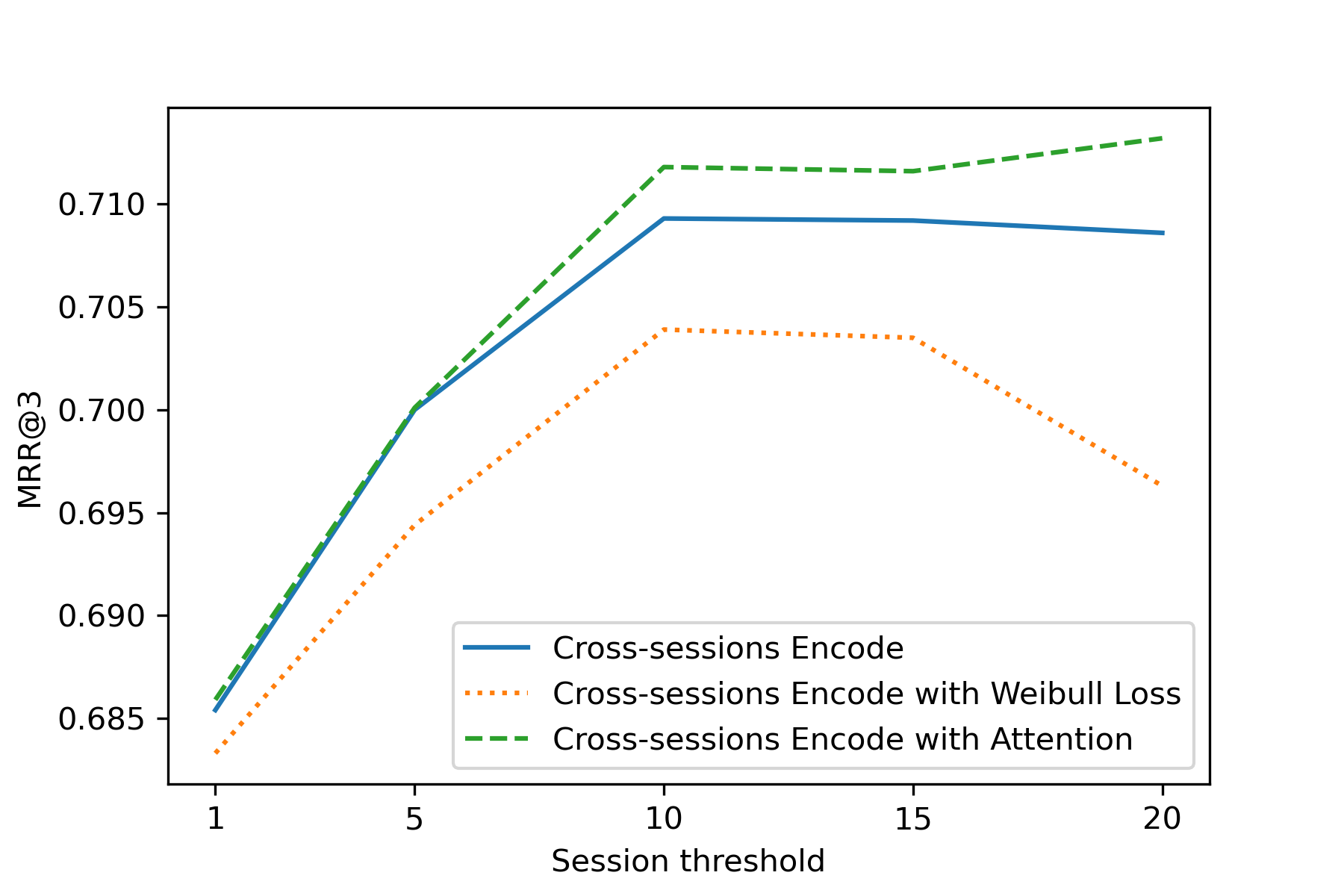}
    \end{subfigure}
    \caption{HR@k and MRR@k for varying choices of the session threshold.}
    \label{fig:session threshold}
\end{figure}

\paragraph{Analysis of Session Threshold}
In Section~\ref{subsec:threshold} we described how we estimated the $10$ days threshold based on the approach in \citet{HalfakerEtAl2015}. This threshold was then used in Section~\ref{subsec:preprocessing} to discard sessions that exceed $10$ days. We now empirically analyse the estimated threshold by varying it, then retrain and evaluate how our models perform with different choices of threshold. The results are presented in Figure~\ref{fig:session threshold}. We observe that the performance of our models drops when decreasing the threshold to less than $10$, while an increase of the threshold has less impact on the performance. It shows that the additional sessions that are added to the input of our models when increasing the threshold above $10$ are not used by the models. Only the model with attention has increasing performance when the threshold exceeds $10$ days. It is likely due to the attention mechanism being able to use the right information from the additional sessions. 

\paragraph{Analysis of Actions}

\begin{table*}[tb]
\caption{Ablation study of actions. Relative change in parentheses.}
\resizebox{\textwidth}{!}{%
\begin{tabular}{@{}llrrrrr@{}}
\toprule
\multicolumn{2}{c}{Model} & \multicolumn{1}{c}{HR@3} & \multicolumn{1}{c}{Precision@3} & \multicolumn{1}{c}{Recall@3} & \multicolumn{1}{c}{MRR@3} & \multicolumn{1}{c}{MAP@3} \\ \midrule
\multirow{10}{*}{Cross-sessions Encode} & all actions & 0.8380 & 0.3030 & 0.8145 & 0.7093 & 0.6923 \\
 & without E-commerce & 0.7526 (-10.19\%) & 0.2698 (-10.95\%) & 0.7249 (-11.00\%) & 0.5951 (-16.09\%) & 0.5764   (-16.74\%) \\
 & without Claims reporting & 0.8250 (-1.55\%) & 0.2979 (-1.68\%) & 0.8012 (-1.64\%) & 0.7006 (-1.22\%) & 0.6829 (-1.35\%) \\
 & without Information & 0.8317 (-0.75\%) & 0.3000 (-0.98\%) & 0.8072 (-0.89\%) & 0.7045 (-0.68\%) & 0.6863 (-0.87\%) \\
 & without Personal account & 0.8067 (-3.73\%) & 0.2894 (-4.48\%) & 0.7803 (-4.19\%) & 0.6604 (-6.89\%) & 0.6438 (-7.00\%) \\
 & without Items & 0.7379 (-11.94\%) & 0.2652 (-12.46\%) & 0.7116 (-12.63\%) & 0.5720 (-19.36\%) & 0.5548 (-19.86\%) \\
 & without Services & 0.8032 (-4.15\%) & 0.2880 (-4.93\%) & 0.7765 (-4.66\%) & 0.6639 (-6.40\%) & 0.6465 (-6.61\%) \\
 & without Start & 0.8162 (-2.60\%) & 0.2935 (-3.11\%) & 0.7906 (-2.94\%) & 0.6771 (-4.55\%) & 0.6592 (-4.77\%) \\
 & without Act & 0.8318 (-0.73\%) & 0.3005 (-0.80\%) & 0.8082 (-0.77\%) & 0.7035 (-0.82\%) & 0.6864 (-0.85\%) \\
 & without Complete & 0.8317 (-0.75\%) & 0.3005 (-0.82\%) & 0.8078 (-0.82\%) & 0.7036 (-0.80\%) & 0.6861 (-0.89\%) \\ \midrule
\multirow{10}{*}{\begin{tabular}[c]{@{}l@{}}Cross-sessions Encode\\ with Weibull Loss\end{tabular}} & all actions & 0.8378 & 0.3013 & 0.8120 & 0.7039 & 0.6852 \\
 & without E-commerce & 0.7433 (-11.28\%) & 0.266 (-11.71\%) & 0.714 (-12.07\%) & 0.5791 (-17.74\%) & 0.5603 (-18.23\%) \\
 & without Claims reporting & 0.8265 (-1.35\%) & 0.2971 (-1.40\%) & 0.8004 (-1.43\%) & 0.6901 (-1.97\%) & 0.6717 (-1.97\%) \\
 & without Information & 0.8329 (-0.59\%) & 0.3006 (-0.22\%) & 0.8081 (-0.48\%) & 0.7013 (-0.37\%) & 0.6835 (-0.25\%) \\
 & without Personal account & 0.8095 (-3.38\%) & 0.2887 (-4.19\%) & 0.7821 (-3.68\%) & 0.6674 (-5.18\%) & 0.65 (-5.14\%) \\
 & without Items & 0.7349 (-12.28\%) & 0.2622 (-12.97\%) & 0.7053 (-13.14\%) & 0.5628 (-20.05\%) & 0.5446 (-20.52\%) \\
 & without Services & 0.8096 (-3.37\%) & 0.2891 (-4.04\%) & 0.7817 (-3.73\%) & 0.6609 (-6.12\%) & 0.6429 (-6.17\%) \\
 & without Start & 0.8149 (-2.73\%) & 0.2909 (-3.46\%) & 0.787 (-3.07\%) & 0.6722 (-4.50\%) & 0.6538 (-4.59\%) \\
 & without Act & 0.8279 (-1.18\%) & 0.2979 (-1.11\%) & 0.8026 (-1.16\%) & 0.6993 (-0.65\%) & 0.6813 (-0.57\%) \\
 & without Complete & 0.8283 (-1.13\%) & 0.2973 (-1.32\%) & 0.8019 (-1.24\%) & 0.6925 (-1.62\%) & 0.674 (-1.63\%) \\ \midrule
\multirow{10}{*}{\begin{tabular}[c]{@{}l@{}}Cross-sessions Encode\\ with Attention\end{tabular}} & all actions & 0.8385 & 0.3026 & 0.8141 & 0.7118 & 0.6941 \\
 & without E-commerce & 0.745   (-11.15\%) & 0.2675 (-11.61\%) & 0.7181 (-11.79\%) & 0.5938 (-16.57\%) & 0.5753 (-17.11\%) \\
 & without Claims reporting & 0.8341 (-0.53\%) & 0.3002 (-0.82\%) & 0.8091 (-0.62\%) & 0.7036 (-1.15\%) & 0.6853 (-1.26\%) \\
 & without Information & 0.8336 (-0.59\%) & 0.3015 (-0.36\%) & 0.81 (-0.50\%) & 0.7085 (-0.46\%) & 0.6914 (-0.38\%) \\
 & without Personal account & 0.8147 (-2.84\%) & 0.2917 (-3.60\%) & 0.7886 (-3.13\%) & 0.676 (-5.02\%) & 0.6592 (-5.03\%) \\
 & without Items & 0.7399 (-11.76\%) & 0.2645 (-12.61\%) & 0.7117 (-12.58\%) & 0.5696 (-19.98\%) & 0.5522 (-20.44\%) \\
 & without Services & 0.8171 (-2.55\%) & 0.2927 (-3.27\%) & 0.7906 (-2.88\%) & 0.6741 (-5.29\%) & 0.6569 (-5.36\%) \\
 & without Start & 0.8194 (-2.28\%) & 0.2953 (-2.44\%) & 0.7942 (-2.44\%) & 0.6831 (-4.03\%) & 0.6651 (-4.17\%) \\
 & without Act & 0.8348 (-0.44\%) & 0.3002 (-0.82\%) & 0.8088 (-0.65\%) & 0.7034 (-1.18\%) & 0.6848 (-1.33\%) \\
 & without Complete & 0.8385 (0\%) & 0.3028 (0.06\%) & 0.8142 (0.02\%) & 0.706 (-0.81\%) & 0.689 (-0.73\%) \\ \bottomrule
\end{tabular}
}%
\label{tab:action_influence}
\end{table*}

We use ablation to analyse the influence of different actions (i.e., sections, objects and types).
Each time, we remove all actions of a given type and evaluate our cross-sessions models after retraining without the action type under analysis.
The results are presented in Table~\ref{tab:action_influence} for all our models and evaluation measures.
We did not consider the action type ``click'' in the ablation study because removing clicks results in removing most of the actions ($65\%$), but also most of the objects and sections since users interact with objects and sections mainly through clicks.

For the Cross-sessions Encode and Cross-sessions Encode with Weibull Loss, all actions contribute positively to the model performance as their removal causes a drop in performance.
We have a similar conclusion for the Cross-sessions Encode with Attention model with the exception that the removal of actions of the type ``complete'' increases performance instead of decreasing it, but the actual difference is negligible (less than $0.5\%$).

Not surprisingly, the most negative impact (up to $-20.52\%$ in MAP) is obtained when actions with the object type ``item'' are removed.
Even if these are not the most frequent objects (see Table~\ref{tab:actions}), they are highly informative as they provide information on the user's interests.
The most frequent objects are ``services'', twice more frequent than items, and their removal affects negatively the performance even if with a less severe impact (up to $-6.61\%$ in MAP).

The second greatest negative impact is obtained when the actions from the section ``e-commerce'' are removed (up to $-18.23\%$ in MAP).
Again, this is not the most frequent section, but it is highly informative since the user needs to access the e-commerce section to inspect and compare different insurance products.
The most frequent section is the ``personal account'', which is $3$ times more frequent than the e-commerce section.
Its removal has a negative impact, but not as severe as for e-commerce (up to $-7\%$ in MAP).

In terms of action type, the removal of the type ``start'' has the greatest negative effect, even if limited with respect to the other two categories (up to $-4.77\%$ in MAP).
The types ``act'' and ``complete'' have a negligible impact (less than $2\%$), but they also represent a very sparse signal (less than $5\%$ of all types together). 

\begin{figure}[tb]
    \centering
    \begin{subfigure}{0.4\textwidth}
        \centering
        \includegraphics[width=\columnwidth]{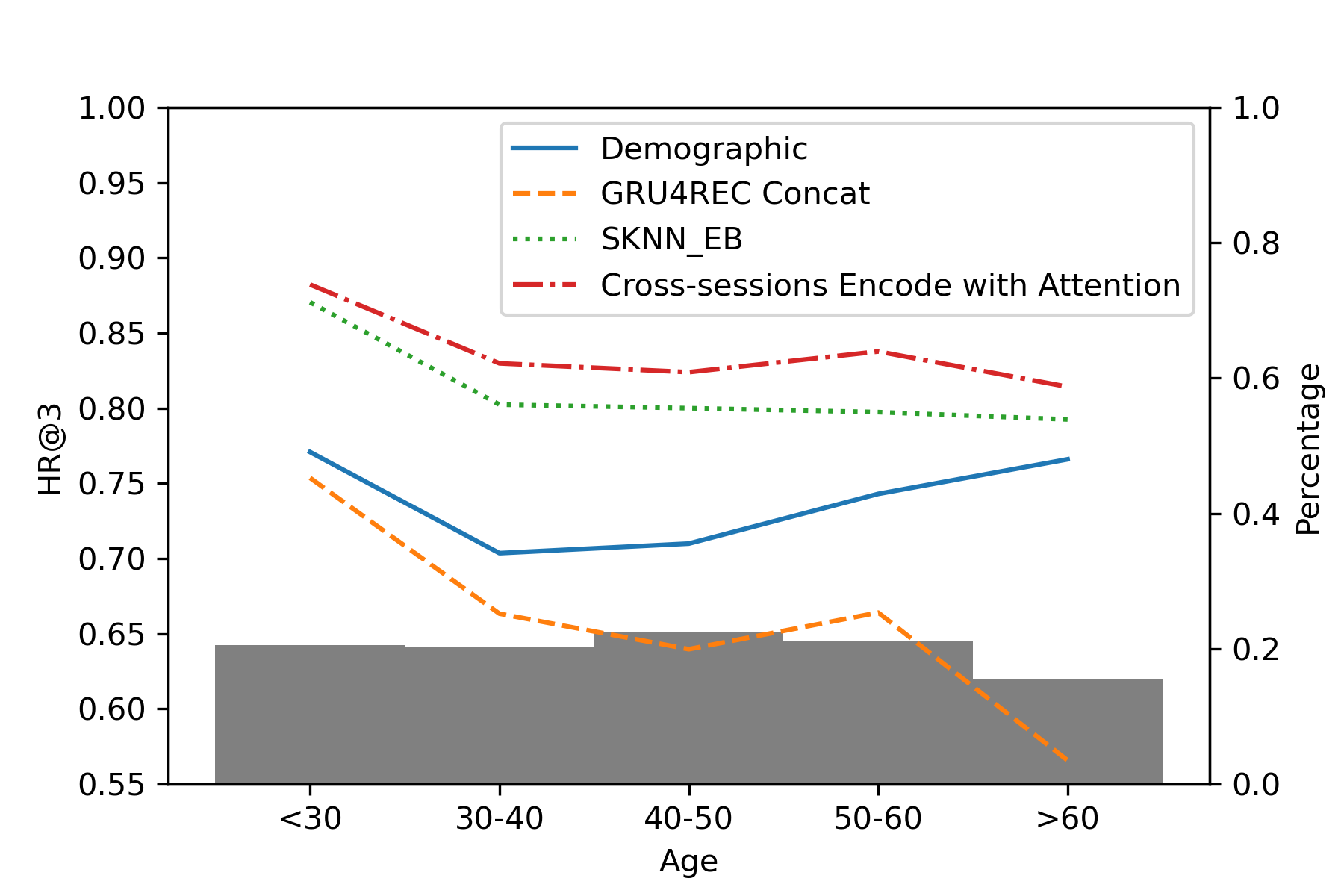}
    \end{subfigure}
    \begin{subfigure}{0.4\textwidth}
        \centering
        \includegraphics[width=\columnwidth]{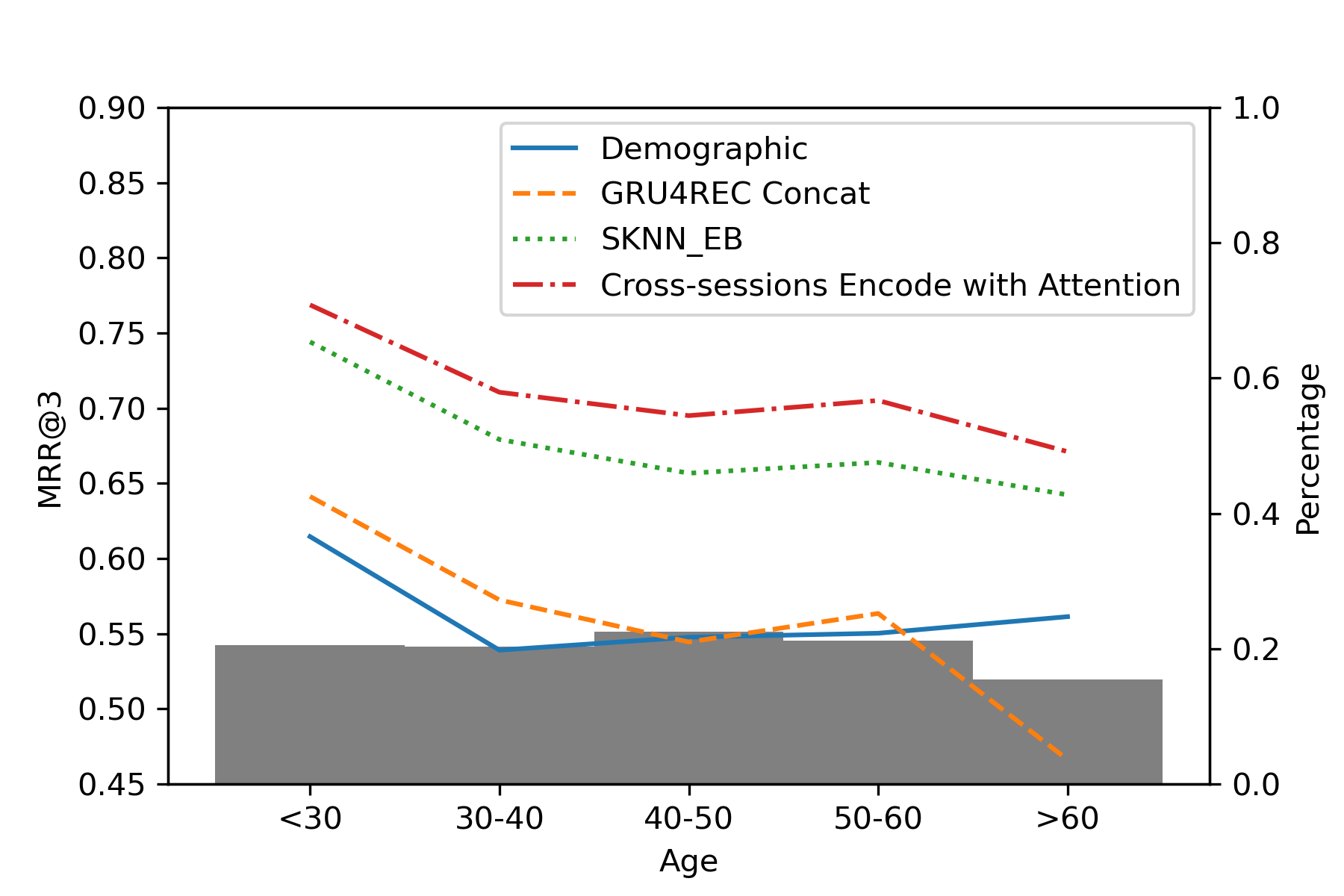}
    \end{subfigure}
    \caption{HR@k and MRR@k for different ages of the users. The grey bars illustrate the percentage of the different groups in the dataset.}
    \label{fig:age}
\end{figure}

\begin{figure}[tb]
    \centering
    \begin{subfigure}{0.4\textwidth}
        \centering
        \includegraphics[width=\columnwidth]{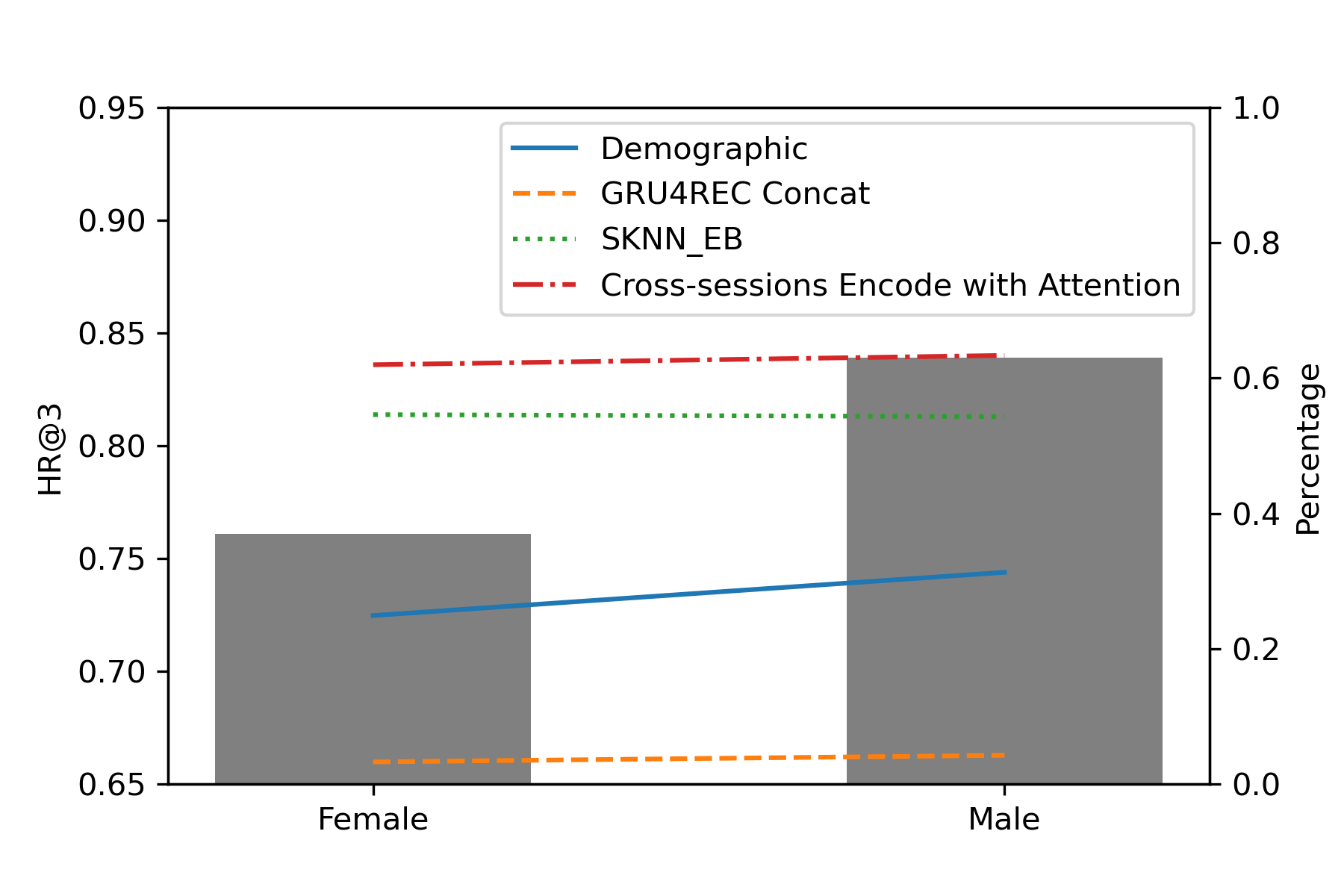}
    \end{subfigure}
    \begin{subfigure}{0.4\textwidth}
        \centering
        \includegraphics[width=\columnwidth]{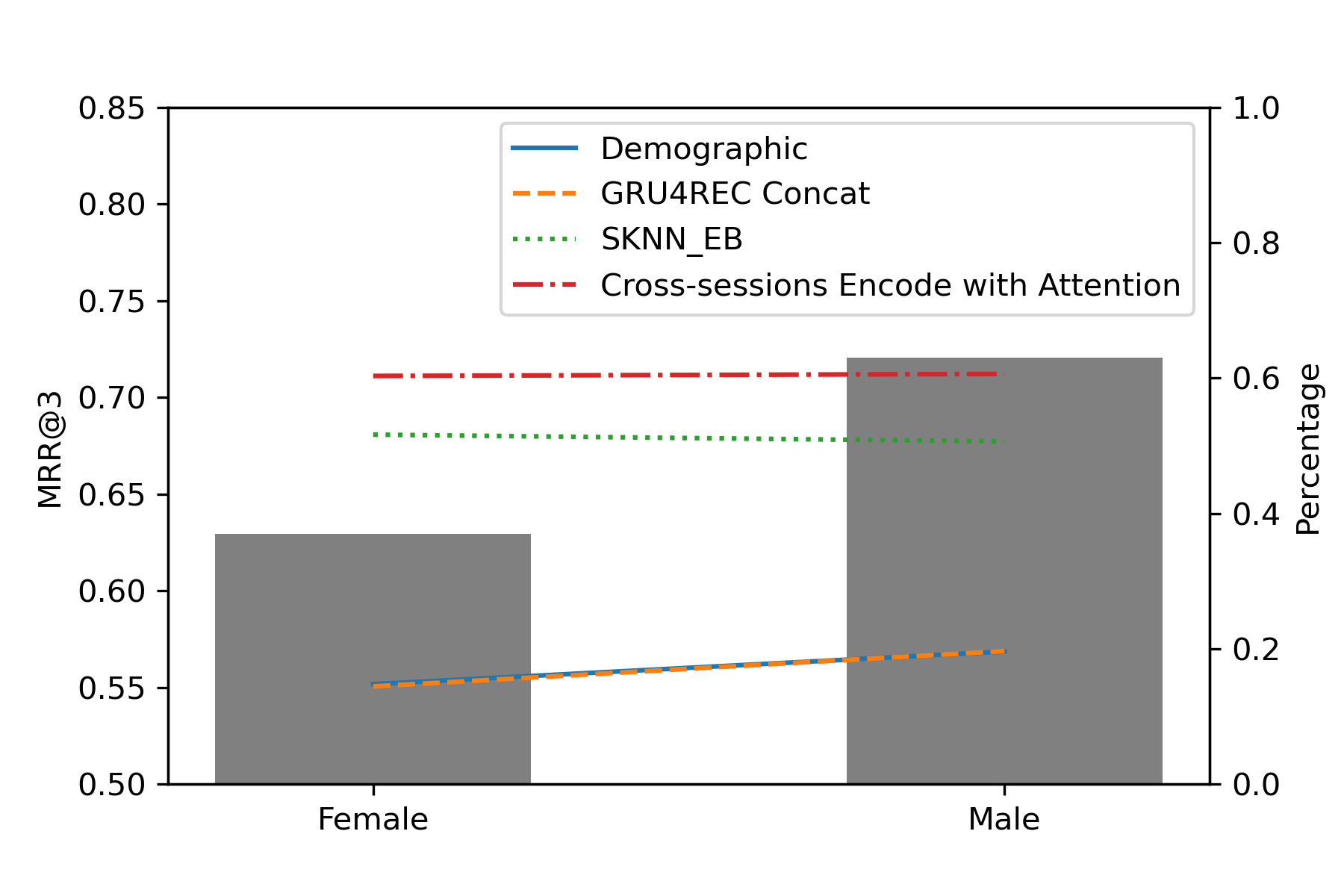}
    \end{subfigure}
    \caption{HR@k and MRR@k for different genders of the users. The grey bars illustrate the percentage of the different groups in the dataset.}
    \label{fig:gender}
\end{figure}

\begin{figure}[tb]
    \centering
    \begin{subfigure}{0.4\textwidth}
        \centering
        \includegraphics[width=\columnwidth]{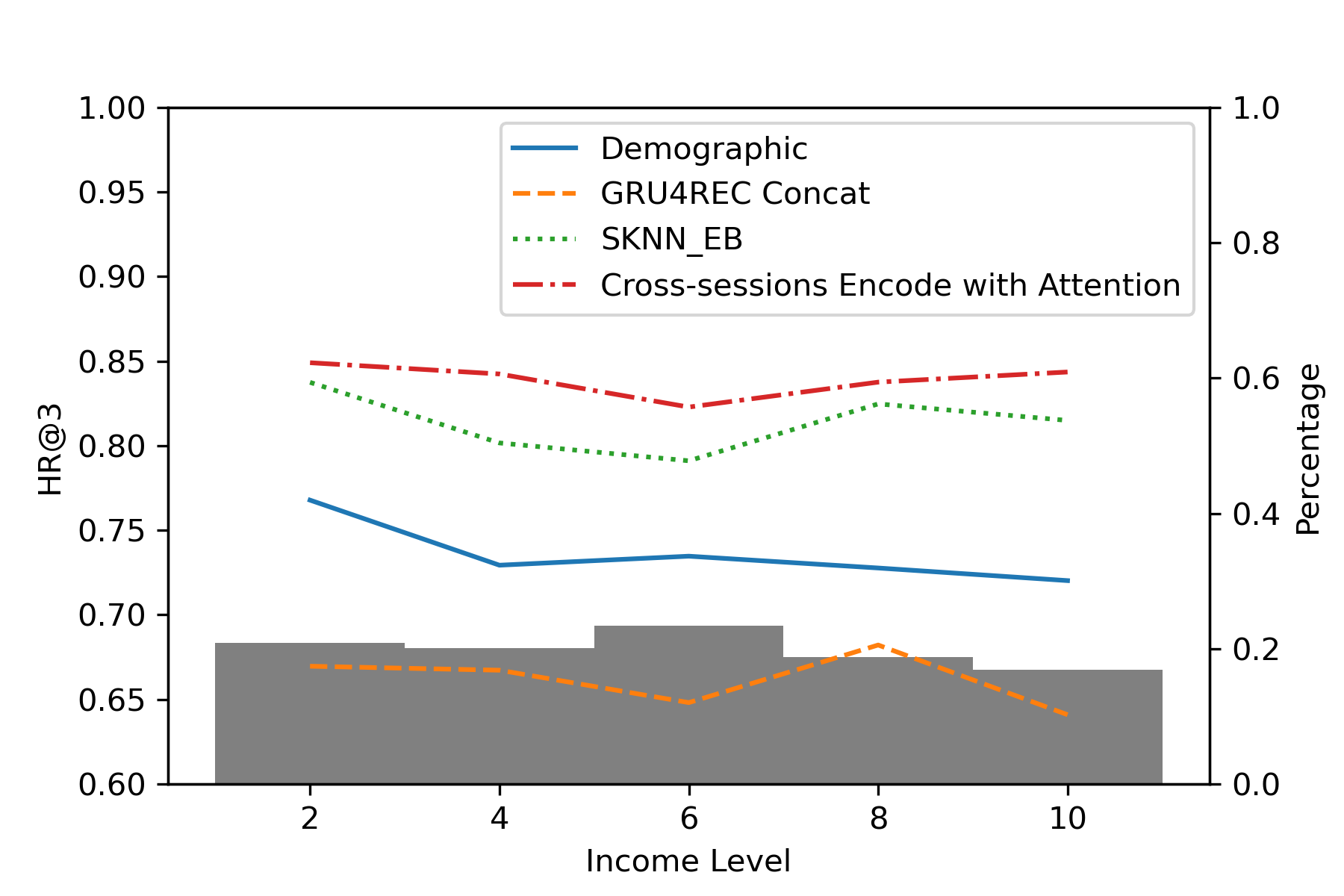}
    \end{subfigure}
    \begin{subfigure}{0.4\textwidth}
        \centering
        \includegraphics[width=\columnwidth]{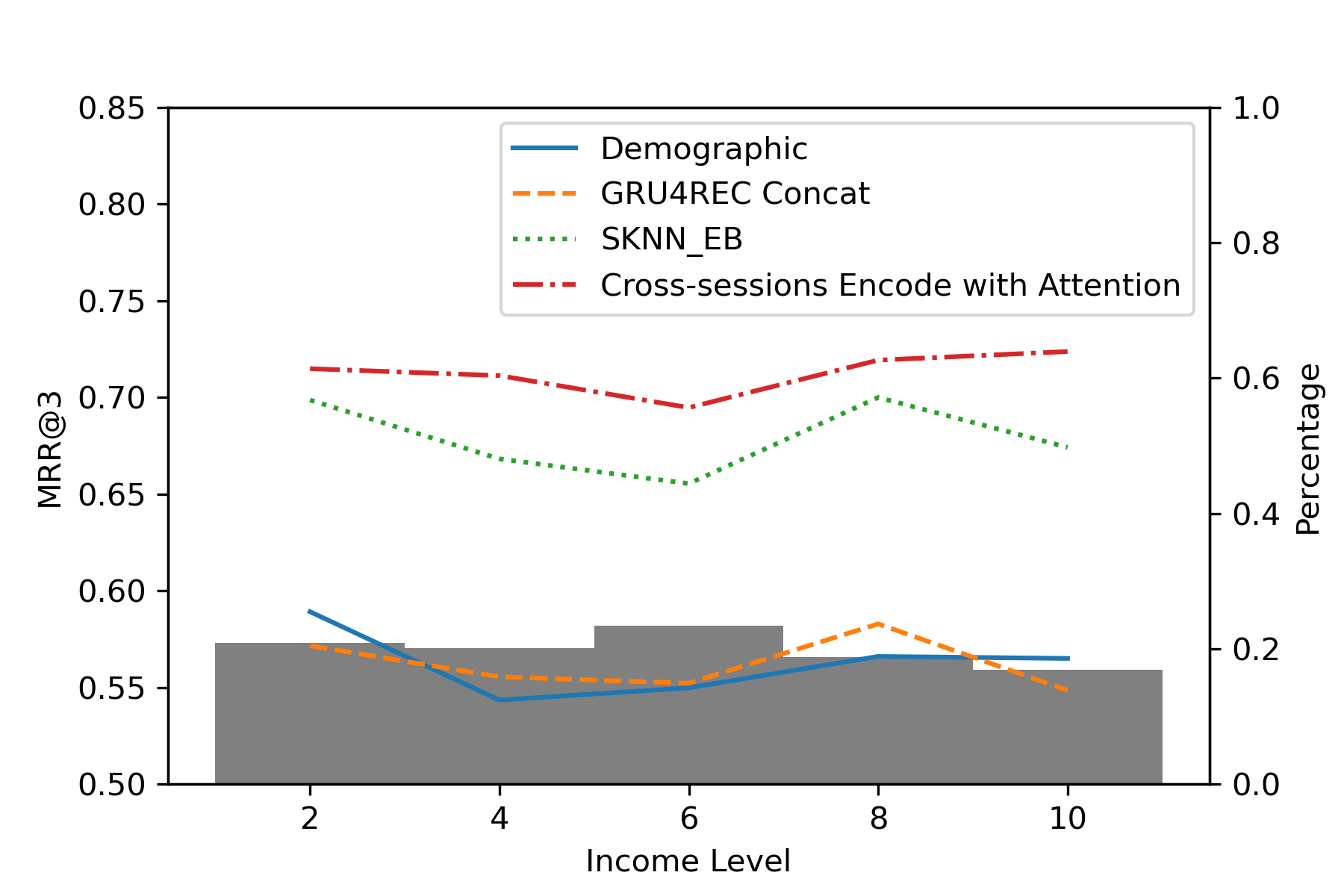}
    \end{subfigure}
    \caption{HR@k and MRR@k for different income levels of the users on a scale from $1-10$, where $1$ is the lowest income level and $10$ is the highest. The grey bars illustrate the percentage of the different groups in the dataset.}
    \label{fig:income}
\end{figure}

\paragraph{Fairness Analysis}
We analyse group fairness of our models compared to the baseline models by exploring the performance for different age, gender and income level of the users, as these are protected characteristics that should not be discriminated against. The results of HR and MRR are illustrated in Figures~\ref{fig:age}, \ref{fig:gender} and \ref{fig:income}. The results for the remaining measures are similar, but not included for brevity. We further explore statistical significance, where Wald test is used for HR and t-test is used for MRR to compare the performance of the same model for different groups. In order to focus on variations between different types of models, we included the best-performing model among the non-sessions-based models, the best-performing version of the GRU4REC model, the best-performing version of the SKNN model and the best-performing model among our cross-sessions models.

We observe that the session-based models (GRU4REC Concat, SKNN\_EB and Cross-sessions Encode with Attention) have lower performance when increasing age of the users. Especially the GRU4REC model has very poor performance for users from the age of $60$, even though this group is well represented in the data. This is also supported by statistical tests, showing that the performance for users younger than $30$ is significantly higher and the performance for users from the age of $60$ is significantly lower than the rest. It does not apply to the same extent for the demographic model.
It indicates that is it more difficult to learn insurance recommendations from user sessions generated by older users.

We further observe that none of the models are unfair towards gender, even though male users are over-represented in the dataset compared to female users. Only the demographic model has slightly better performance for male users than female users. Statistical tests show that none of the models have significantly different performances between genders.

Finally, we do not observe any strong trends in the performance for different income levels of the users. The demographic model has slightly decreasing performance for users with higher income level (only income level $1-2$ is statistically different from the others), while all the session-based models have slightly lower performance for users with middle-income level (it is only statistically significant for the SKNN model).

\section{Conclusions and Future Work}
\label{sec:conclusions}

The insurance domain is a data-scarce domain, where there is a limited amount of user feedback, interactions and number of items.
These features make insurance recommendations particularly challenging.
We have addressed the problem of insurance recommendation with our cross-sessions models, which learn to recommend items from past user sessions and different types of user actions.
These actions are not always directly associated with items, for example a user reporting a claim on the insurance website.
Unlike state-of-the-art session-based and session-aware models, our models predict a target action that does not occur within the session.
Our cross-sessions models are all based on RNNs with GRU units.
These are combined with $3$ different ways to represent user sessions: maximum pooling encoding, concatenation of multiple sessions and an autoencoder approach; and $3$ different loss functions and architectures: cross-entropy, censored Weibull and attention mechanism.

Experimental results on a real-world dataset show that all our cross-sessions models outperform several state-of-the-art baselines.
Moreover, combining our models with demographic features boosts the performance even further.
Additional analyses on the results shows that: (1) considering past user sessions is beneficial for cross-sessions models; (2) the removal of any type of action harms the performance, thus considering several types of actions is also beneficial.
Finally, an analysis of group fairness with respect to age, gender, and income level, shows that our cross-session models are not biased against specific groups.

The output vector $o$, which was introduced in Section \ref{subsec:approach}, contains a score for each item. The length of the vector is thereby equal to the total number of items. In item-poor domains like the insurance domain, the size of vector $o$ is consequently small, but in other domains like music and retail, it is not rare to have hundreds of thousands of items. In such a case, calculating a score for each item would not be the most efficient approach. A possible solution could be to sample the output and only compute the score for a small subset of the items together with some negative examples. In future work, the effect of the size of vector $o$ in our proposed models could be studied in large item domains with different sampling strategies.

As future work, we further plan to run an online A/B testing experiment to evaluate our models in an online scenario with users interacting in real-time.
Moreover, we will investigate how to combine ML interpretability models~\cite{Molnar2020}, for example Local Surrogate (LIME), SHapley Additive exPlanations (SHAP), with our cross-sessions models to generate automatic explanations that will be shown to online customers.
Finally, we will combine our models with other types of data, for example transcripts of user conversations over the phone, and develop a multi-modal extension of our models.



\bibliographystyle{ACM-Reference-Format}
\bibliography{sample-base.bib}


\end{document}